\documentclass[a4paper,11pt]{article}
\pdfoutput=1 % if your are submitting a pdflatex (i.e. if you have
\usepackage{jheppub} % for details on the use of the package, please
\usepackage[T1]{fontenc} % if needed
\usepackage{graphicx}
\usepackage{amsmath}
\usepackage{amssymb}
\usepackage[dvipsnames]{xcolor}
\usepackage{centernot}

\def\U1mt{U(1)_{L_\mu-L_\tau}}
\def\wt{\widetilde}
\def\ol{\overline}
\def\nl{\nonumber\\}

\title{\boldmath Dark matter for $b\to s \mu^+ \mu^-$ anomaly in a gauged $U(1)_X$ model}

\author{Seungwon Baek,}
\emailAdd{sbaek@korea.ac.kr}
\affiliation{Department of Physics, Korea University, \\
Anam-ro 145, Sungbuk-gu, Seoul 02841, Korea}

 \author{Chaehyun Yu}
\emailAdd{chyu@korea.ac.kr}

\abstract{We propose a new physics model which has a cold dark matter candidate and can explain the $b \to s \mu^+\mu^-$ anomaly at the same
time. Our model includes  a  scalar quark $\widetilde{q}$ and a scalar lepton $\widetilde{\ell}$ which are  $SU(2)_L$-doublet as
well as a Dirac fermion $N$ which is $SU(2)_L$-singlet. The new particles are charged under a gauged $U(1)_X$ group which is
spontaneously broken to a discrete $Z_2$ symmetry by a dark scalar $S$. The remnant $Z_2$ symmetry stabilizes the dark matter.
Box diagrams with $\widetilde{q}$, $\widetilde{\ell}$, and $N$ running inside the loop can generate the correct Wilson coefficients
$C_9^\mu = -C_{10}^\mu$ to accommodate the $b \to s \mu^+\mu^-$ anomaly while avoiding constraints such as $B_s-\ol{B}_s$ mixing.
The dark matter annihilation into a second generation lepton pair via $t$-channel $\widetilde{\ell}$-exchanging process plays an important
role in producing the current dark matter relic abundance of the universe, showing a strong interplay between the flavor and dark matter physics.
We also discuss dark-gauge-interaction-dominated and Higgs-portal-dominated
scenarios for dark matter physics.
}

\begin{document} 
\maketitle
\flushbottom

\section{Introduction}
\label{sec:intro}

Since the LHCb Collaboration reported some deviations from the Standard Model (SM) prediction in the $B\to K^{(*)}\ell \ell$ ($\ell=e, \mu$)
decays a few years ago~\cite{Aaij:2014ora,Aaij:2013qta}, a lot of interest has been drawn to reveal the origin of the anomalies~\cite{
Chang:2013hba, Crivellin:2015lwa,Sierra:2015fma,Allanach:2015gkd,
  Boucenna:2016wpr,Boucenna:2016qad,Kawamura:2017ecz,
  GarciaGarcia:2016nvr,Ko:2017quv, Ko:2017yrd, Ko:2017lzd, King:2017anf,
  DiChiara:2017cjq,Alonso:2017bff,Bonilla:2017lsq,Ellis:2017nrp,
Alonso:2017uky,Tang:2017gkz,Datta:2017ezo,Chiang:2017hlj,Choudhury:2017qyt,
Bauer:2015knc,Das:2016vkr,Becirevic:2016yqi,Sahoo:2016pet,Hiller:2016kry,Bhattacharya:2016mcc,Becirevic:2017jtw,
Cai:2017wry,Das:2017kfo,
Belanger:2015nma,Gripaios:2015gra,Hu:2016gpe,Crivellin:2017zlb,DAmico:2017mtc,Celis:2017doq,Kamenik:2017tnu,Poh:2017tfo,Cline:2017ihf,
Calibbi:2017qbu,Dalchenko:2017shg,Cline:2017aed,Faisel:2017glo,Chiang:2017zkh,
Bian:2017xzg,Cline:2017qqu,Choudhury:2017ijp,DiLuzio:2017fdq,Fuyuto:2017sys,Blanke:2018sro,Dasgupta:2018nzt,Iguro:2018qzf,Chala:2018igk,Falkowski:2018dsl,Vicente:2018xbv,Mu:2018weh,Liu:2018xsw}.
The relevant process to the anomalies in the quark level is the $b\to s \ell \ell$ transition, which is flavor-changing neutral current
(FCNC) and is highly suppressed in the SM. Therefore, the semileptonic $B$ decays would greatly be sensitive to new physics (NP).

The sizable discrepancies reported by the LHCb Collaboration are the ratio, $R_{K^{(*)}}$, of branching ratios
of the $B$ decays into $K^{(*)}\ell \ell$, which is defined by
\begin{equation}
R_{K^{(*)}}=\frac{{\cal B}(B\to K^{(*)} \mu^+ \mu^-)}{{\cal B}(B\to K^{(*)} e^+ e^-)}
\end{equation}
with the SM prediction close to unity.
$R_K$ for the dilepton invariant mass squared range $1<q^2<6$ GeV$^2$  in the $B^+\to K^+ \ell^+ \ell^-$ decay has $2.6\sigma$ deviation 
from the SM prediction~\cite{Aaij:2014ora}, 
while the $R_{K^*}$ values are deviated from the SM predictions
by $2.1$--$2.3\sigma$ and $2.4$--$2.5\sigma$ in the low ($0.045 < q^2 < 1.1$ GeV$^2$) and high ($1.1<q^2<6$ GeV$^2$) dilepton invariant mass region, respectively~\cite{Aaij:2017vbb}. 
Another anomaly in the $b\to s \ell \ell$ transition reported by the LHCb Collaboration is the differential branching fraction for $1<q^2<6$ GeV$^2$ in the $B_s \to \phi \mu^+\mu^-$ decay, which is more than $3\sigma$ below the SM predictions based on the light-cone sum-rule form factors~\cite{Aaij:2015esa}.
Finally, the angular analyses of $B \to K^{*} \ell^+ \ell^-$ performed 
by the LHCb, BELLE, and ATLAS collaborations show about $2 \sim 3\sigma$ deviation 
for the $P_5^\prime$ observable~\cite{Aaij:2013qta,Aaij:2015oid,Wehle:2016yoi,ATLAS:2017dlm} 
while the measurement by the CMS Collaboration is consistent with the SM~\cite{Sirunyan:2017dhj}.
These observations may imply hints of NP in the $b\to s\ell\ell$ transition. 
The $b \to s \ell \ell$ transition is described by the effective weak Hamiltonian
\begin{align}
{\cal H}_{\rm eff} &= -{4 G_F \over \sqrt{2}} V_{ts}^* V_{tb} \sum_i (C_i^\ell O_i^\ell +
                     C_i^{\prime \ell} O_i^{\prime \ell})+h.c.,
\end{align}
where 
\begin{align}
O_{7\gamma}^{(\prime)} &= {e \over 16 \pi^2} m_b (\bar{s} \sigma^{\mu\nu} P_{R(L)} b) F_{\mu\nu}, \quad
O_{8g}^{(\prime)} = {g_s \over 16 \pi^2} m_b (\bar{s} \sigma^{\mu\nu}  T^a P_{R(L)} b) G^a_{\mu\nu}, \nl
O_9^{(\prime)\ell} &= {e^2 \over 16 \pi^2} (\bar{s} \gamma_\mu P_{L(R)} b)(\bar{\ell} \gamma^\mu \ell), \quad
O^{(\prime)\ell}_{10} = {e^2 \over 16 \pi^2} (\bar{s} \gamma_\mu P_{L(R)} b)(\bar{\ell} \gamma^\mu \gamma_5 \ell), \nl
O_S^{(\prime)\ell} &= {e^2 \over 16 \pi^2} m_b (\bar{s}  P_{R(L)} b)(\bar{\ell} \ell), \quad
O_P^{(\prime)\ell} = {e^2 \over 16 \pi^2} m_b (\bar{s}  P_{R(L)} b)(\bar{\ell} \gamma_5 \ell).
\label{eq:effective}
\end{align}
Writing $C_i^{(\prime)\ell} = C_i^{(\prime)\rm SM}+C_i^{(\prime)\ell,\rm NP}$, we obtain $C_{7\gamma}^{\rm SM} \simeq -0.294$,
$C_9^{\rm SM}\simeq 4.20$, $C_{10}^{\rm SM}\simeq -4.01$ at $m_b$ scale~\cite{Bobeth:1999mk,Mahmoudi:2018qsk}. 
Global fitting analyses~\cite{Descotes-Genon:2013wba,Altmannshofer:2017fio, Capdevila:2017bsm,
  Alok:2017jaf, Ciuchini:2017mik,Alok:2017sui} show that sizable NP contributions to $C_{9 (10)}^\mu$ can accommodate the data.
We notice that the individual deviations in the observables mentioned above are in the same
direction,  {\it i.e.} destructive with the  SM, and when combined, the discrepancy with the
SM predictions can be as large as $\sim5\sigma$~\cite{Descotes-Genon:2013wba,Altmannshofer:2017fio, Capdevila:2017bsm,
  Alok:2017jaf, Ciuchini:2017mik,Alok:2017sui}.
Best fit values of NP models manifesting in a  one-dimensional Wilson coefficient(s) include $C_9^{\mu,{\rm NP}}=-1.21 (5.2 \sigma)$ and 
$C_9^{\mu,{\rm NP}}=-C_{10}^{\mu,{\rm NP}}=-0.67 (4.8 \sigma)$~\cite{Altmannshofer:2017fio}.
We consider the latter scenario in this paper. The allowed range in this model is
\begin{align}
%-0.81 \le C_9^{\mu,{\rm NP}}&=-C_{10}^{\mu,{\rm NP}} \le -0.51 \;\; (1\sigma). 
-0.97 \le C_9^{\mu,{\rm NP}}&=-C_{10}^{\mu,{\rm NP}} \le -0.37 \;\; (2\sigma). 
\label{eq:C9_range}
\end{align}

On the other hand, other observables relevant to the %$b\to s\ell\ell$ transition are well consistent with the SM.
$b\to s$ transition are well consistent with the SM.
For example, the branching fractions of the pure leptonic decay, $B_s\to \mu^+\mu^-$, and the radiative decay, $B\to X_s\gamma$, agree with
the SM estimations. In addition, the forward-backward asymmetry, $A_{FB}$,
and the quantity, $F_H$, which are defined in the $B^+\to K^+\mu^+\mu^-$ decay as $d\Gamma/d\cos\theta\sim \tfrac{3}{4}(1-F_H)(1-\cos^2\theta)
+\tfrac{1}{2}F_H + A_{FB}\cos\theta$,
are good agreement with the SM prediction~\cite{Bobeth:2007dw}. 
Especially, the latter disfavors the presence of new (pseudo)scalar operators, $O_S^{(\prime)\ell}$, $O_P^{(\prime)\ell}$, 
or tensor operators, $O_{7\gamma}^{(\prime)}$, $O_{8g}^{(\prime)}$,  in the $b\to s\mu^+\mu^-$ decay~\cite{Descotes-Genon:2015uva}.

In the present article, we propose a NP model with an additional local dark $U(1)_X$ symmetry to resolve the anomalies in the $b\to s
\ell\ell$ transition. We introduce a new vector-like fermion and several new scalars charged under the $U(1)_X$ symmetry as well as the SM
gauge symmetry. The SM particles are neutral under the $U(1)_X$ symmetry. The new fermion and scalars can interact with the SM fermions
through Yukawa interactions and, also, the mixing between the SM Higgs doublet and new scalar singlet. Because the gauge boson of the
$U(1)_X$ symmetry does not couple to the SM fermions directly, the $b\to s\ell\ell$ transition can have effects of NP through box diagrams at
the loop level. We find that this model could account for the anomalies in the $b\to s \ell \ell$ transition. This model naturally contains a
candidate for cold dark matter (DM) due to the remnant $Z_2$ symmetry after breakdown of the $U(1)_X$ symmetry. We find that this model can
also explain the relic density of the universe.

This paper is organized as follows. In section~\ref{sec:model}, we construct our model. In section~\ref{sec:NP} we calculate NP
contribution to $b \to s \mu\mu$, $B \to K^{(*)} \nu \ol{\nu}$, $B_s - \ol{B}_s$ mixing, $B_s\to \mu^+\mu^-$, $b \to s \gamma$, the anomalous magnetic moment of
muon $a_\mu$, and the loop-induced effective $Z \mu^+ \mu^-$ coupling. In section~\ref{sec:DM} we consider dark matter phenomenology.
Finally we conclude in section~\ref{sec:concl}. Loop functions are collected in appendix~\ref{app:loop}.

\section{The model}
\label{sec:model}
In addition to the SM gauge group $SU(3)_C \times SU(2)_L \times U(1)_Y$ we introduce a local dark $U(1)_X$ symmetry under which all the SM
fields are neutral.
We also introduce new fields 
\begin{align}
N, 
\quad \wt{q}\equiv \left(\begin{array}{c} \wt{u} \\ \wt{d} \end{array}\right),
\quad \wt{\ell}\equiv \left(\begin{array}{c} \wt{\nu} \\ \wt{e} \end{array}\right),
\quad S,
\end{align}
which have quantum number assignments as shown in table~\ref{tab:particles}.

\begin{table}[th]
\begin{center}
\begin{tabular}{|c|c|c|c|c|}\hline
&\multicolumn{1}{|c|}{New fermion} & \multicolumn{3}{c|}{New scalars} \\\hline
                 & $N$             & $\wt{q}$   & $\wt{\ell}$    & $S$ \\ \hline
 $SU(3)_C$ &  {\bf 1}         &  {\bf 3}        & {\bf 1}    &  {\bf 1}   \\ \hline
 $SU(2)_L$ &  {\bf 1}         &  {\bf 2}          & {\bf 2}  &  {\bf 1}   \\ \hline
 $U(1)_Y$ &  $0$              &  ${1 \over 6}$      & $ -{1 \over 2}$      &  $0$   \\ \hline
 $U(1)_X$ & $Q$ & $-Q$  & $-Q$ & $2Q$     \\\hline
\end{tabular}
\caption{Assignments of quantum numbers for $N, \wt{q}, \wt{\ell}$ and $S$ under the gauge group $SU(3)_C \times SU(2)_L \times U(1)_Y \times U(1)_X$. }
\label{tab:particles}
\end{center}
\end{table}

The Dirac fermion $N$ has a mass term,
\begin{align}
{\cal L}_{\rm mass} &= -M_N \ol{N} N.
\end{align}
It couples to the $SU(2)_L$-doublet scalars $\wt{q}$, $\wt{\ell}$, and the SM-singlet scalar $S$, via Yukawa interactions,
\begin{align}
{\cal L}_{\rm Yukawa} &= - \sum_{i=1,2,3} \lambda_q^i \ol{q}_L^i \wt{q} N - \sum_{i=1,2,3} \lambda_\ell^i \ol{\ell}_L^i \wt{\ell} N   -{f \over 2} \ol{N^c} N S^\dagger + h.c.,
\label{eq:Lag_Y}
\end{align}
where $i(=1,2,3)$ is the generation index. We set $\lambda_q^1 \equiv 0$ to evade strong constraints, {\it e.g.}, from $B_d^0-\ol{B}_d^0$
mixing. 
The Yukawa couplings  $\lambda_\ell^1$ and $\lambda_\ell^3$ are irrelevant to $b\to s \mu\mu$ transition, and we set
$\lambda_\ell^1 \equiv \lambda_\ell^3 \equiv 0$ in order not to generate $\mu \to e \gamma$ and $\tau \to \mu (e) \gamma$ processes.
The scalar potential is written in the form
\begin{align}
V &= V(H,S) + V(H, S, \wt{q}, \wt{\ell}).
\label{eq:V}
\end{align}
Here $V(H,S)$ has terms involving the SM Higgs doublet $H$ and the singlet $S$ which get non-vanishing vacuum expectation values  (VEVs), 
$v_H (=\sqrt{2} \langle H^0 \rangle)$ and $v_S (=\sqrt{2} \langle S \rangle)$, as
\begin{align}
V(H,S) &=\lambda_H \left(H^\dagger H-\frac{v_H^2}{2} \right)^2 + \lambda_S \left(S^\dagger S-\frac{v_S^2}{2} \right)^2 + \lambda_{HS}
 \left(H^\dagger H -\frac{v_H^2}{2}\right)\left(S^\dagger S -\frac{v_S^2}{2}\right).
\end{align}
The terms in $V(H,S,\wt{q},\wt{\ell})$ include fields $\wt{q}$ and $\wt{\ell}$ additionally,
\begin{align}
V(H,S,\wt{q},\wt{\ell})&=m^2_{\wt{q}} \wt{q}^\dagger \wt{q} +m^2_{\wt{\ell}} \wt{\ell}^\dagger \wt{\ell} + \lambda_{\wt{q}} \left(\wt{q}^\dagger \wt{q}\right)^2 + \lambda_{\wt{\ell}} \left(\wt{\ell}^\dagger \wt{\ell}\right)^2\nl
&+\lambda_{H\wt{q}} \left(H^\dagger H -\frac{v_H^2}{2}\right) \wt{q}^\dagger \wt{q}  
+  \lambda'_{H\wt{q}} \left(H^\dagger \wt{q}\right)  \left( \wt{q}^\dagger H \right) 
+  \lambda^{\prime\prime}_{H\wt{q}} \left(\wt{H}^\dagger \wt{q}\right) \left( \wt{q}^\dagger \wt{H} \right)\nl
&+\lambda_{H\wt{\ell}} \left(H^\dagger H -\frac{v_H^2}{2}\right) \wt{\ell}^\dagger \wt{\ell}  
+  \lambda'_{H\wt{\ell}} \left(H^\dagger \wt{\ell}\right) \left( \wt{\ell}^\dagger H \right)
+  \lambda^{\prime\prime}_{H\wt{\ell}} \left(\wt{H}^\dagger \wt{\ell}\right) \left( \wt{\ell}^\dagger \wt{H} \right)\nl
&+\lambda_{S\wt{q}} \left(S^\dagger S -\frac{v_S^2}{2}\right) \wt{q}^\dagger \wt{q} +\lambda_{S\wt{\ell}} \left(S^\dagger S -\frac{v_S^2}{2}\right) \wt{\ell}^\dagger \wt{\ell},
\end{align}
where $\wt{H} \equiv i \sigma^2 H^*$.

Now let's consider the particle spectra.  
After $S$ gets VEV, the $U(1)_X$ gauge boson becomes massive with mass,
\begin{align}
m_{Z'} = 2 g_X |Q| v_S, 
\end{align}
where $g_X$ is the gauge coupling constant of $U(1)_X$ group. 
For the dark fermion sector, after diagonalizing the mass matrix 
\begin{align}
\left(
\begin{array}{cc}
M_N & \frac{f v_S}{\sqrt{2}} \\
\frac{f v_S}{\sqrt{2}} & M_N
\end{array}
\right),
\end{align}
obtained in  $(N,N^c)$ basis, we get the mass eigenstates
\begin{align}
N_- &= {1 \over \sqrt{2}} (N- N^c), \nl
N_+ &= {1 \over \sqrt{2}} (N+ N^c),
\label{eq:N_mass}
\end{align}
with masses $M_\mp =M_N \mp f v_S/\sqrt{2}$. From (\ref{eq:N_mass}) we can see that the Majorana phases, $\eta_\mp=\mp 1$, 
are assigned so that $N_\mp^c = \eta_\mp N_\mp$.  
We see that the original Dirac particle $N$ splits into two Majorana fermions $N_\mp$.
It is noted that the VEV $v_S$ breaks the original $U(1)_X$ symmetry into a remnant discrete $Z_2$ symmetry under which $N_\mp$, $\wt{q}$, and $\wt{\ell}$ 
are odd while all the others are even.
By this local discrete symmetry the lightest new particle which we take to be $N_-$ with odd parity under the $Z_2$ symmetry does not decay into
any other particles and can play the
role of a dark matter candidate. 

We can write the SM Higgs $H$ and the dark scalar $S$ in terms of their components
\begin{align}
H =\left( \begin{array}{c} 0 \\ {1 \over \sqrt{2}}(v_H +h) \end{array}\right), \quad
S = {1 \over \sqrt{2}} (v_S + s),
\end{align}
in the unitary gauge. The potential $V$ given in the form of (\ref{eq:V}) automatically satisfies the tadpole condition,
$\partial V/\partial h|_{\rm vacuum} =\partial V/\partial s|_{\rm vacuum} =0$. The scalar mass-squared matrix is obtained 
\begin{equation}
\left(
\begin{array}{cc}
2 \lambda_H v_H^2 & \lambda_{HS} v_H v_S \\
\lambda_{HS} v_H v_S & 2 \lambda_S v_S^2 \\
\end{array}
\right),
\end{equation}
in the basis of $(h,s)$. The above matrix can be diagonalized by introducing mixing angle $\alpha_H$ to get the scalar mass eigenstates $(H_1,H_2)$
\begin{align} 
\left(
\begin{array}{c}
h \\ s
\end{array}
\right)
=
\left(
\begin{array}{cc}
\cos\alpha_H & \sin\alpha_H\\
-\sin\alpha_H & \cos\alpha_H \\
\end{array}
\right)
\left(
\begin{array}{c}
H_1 \\ H_2
\end{array}
\right),
\end{align} 
where $H_1$ is identified with the SM-like Higgs boson with mass $m_{H_1}=125$ GeV. The mixing angle $\alpha_H$ is
constrained by the LHC Higgs experiments~\cite{Baek:2011aa, Baek:2012se}. 
The direct detection experiments of dark matter also constrains this angle through the {\it Higgs portal interaction}, 
$\lambda_{HS} H^\dagger H S^\dagger S$. We take $\alpha_H \le 0.1$ in order to avoid these constraints.

The quark fields $q_L = (u_L, d_L)^T,$\footnote{The generation index is suppressed.} need to be rotated to be in the mass eigenstates. We  assume that the down-type quarks 
in (\ref{eq:Lag_Y}) are already in the mass basis and that the flavor mixing due to 
Cabibbo-Kobayashi-Maskawa (CKM) matrix $V$ appears in the up-quark sector, {\it i.e.} $d_L=d'_L, u_L =V^\dagger u'_L$ 
with $d'_L,u'_L$ being the mass eigenstates. In the mass-eigenstate basis, the Yukawa interactions with quarks are 
\begin{align}
\Delta {\cal L}_{\rm Yukawa} &= -\sum_{i=1,2,3} \left(\lambda_u^i \ol{u}^{\prime i}_L \wt{u} N +\lambda_d^i \ol{d}^{\prime i}_L \wt{d} N \right)+h.c.,
\end{align}
where $\lambda_u^i = \sum_{j=1,2,3} V_{ij} \lambda_q^j$ and $\lambda_d^i = \lambda_q^i (i=1,2,3)$. As a consequence we can see that
\begin{align}
\lambda_u^1 = V_{us} \lambda_q^2 + V_{ub} \lambda_q^3,
\label{eq:lambda_u1}
\end{align}
is induced even if we set $\lambda_q^1 \equiv 0$. The induced $\lambda_u^1$ can be constrained, {\it e.g.}, by $D^0-\ol{D}^0$ mixing.
However, due to Cabibbo-suppressed contribution to $D^0-\ol{D}^0$ at least by  ${\cal O}(\lambda_C^2)$ with $\lambda_C(\approx 0.23)$
  being the Cabibbo angle, the constraint from $D^0-\ol{D}^0$ can be always
satisfied once the constraint from $B_s-\ol{B}_s$ is imposed~\cite{Arnan:2016cpy}. We do not consider this constraint further.
The effective Yukawa coupling $\lambda_\nu^i$ to the $i$-th neutrino is obtained by a similar procedure with the quark case: 
$\lambda_e^i =\lambda_\ell^i$, 
$\lambda_\nu^i = U^*_{ji} \lambda_\ell^j$, where $U$ is the Pontecorvo-Maki-Nakagawa-Sakata (PMNS) mixing matrix
for neutrino oscillations. 

In the new charged-scalar sector, there is mass splitting between $\wt{u} (\wt{\nu})$ and $\wt{d} (\wt{e})$ due to 
$\lambda^{\prime(\prime\prime)}_{H\wt{q}}(\lambda^{\prime(\prime\prime)}_{H\wt{\ell}})$ term:
\begin{align}
m^2_{\wt{u}} = m^2_{\wt{q}}+ {1 \over 2} \lambda^{\prime\prime}_{H\wt{q}} v_H^2, 
& \quad m^2_{\wt{\nu}} = m^2_{\wt{\ell}}+ {1 \over 2} \lambda^{\prime\prime}_{H\wt{\ell}} v_H^2\nl
m^2_{\wt{d}} = m^2_{\wt{q}}+ {1 \over 2} \lambda'_{H\wt{q}} v_H^2, 
& \quad m^2_{\wt{e}} = m^2_{\wt{\ell}}+ {1 \over 2} \lambda'_{H\wt{\ell}} v_H^2.
\end{align}
Since large scalar mass splitting leads large contribution to $\rho$-parameter~\cite{Patrignani:2016xqp}
and also the mass splitting does not affect our analysis, we set 
$\lambda^{\prime(\prime\prime)}_{H\wt{q}}=\lambda^{\prime(\prime\prime)}_{H\wt{\ell}}=0$ for simplicity.

The kinetic mixing term between $Z'$ and $Z(\gamma)$, $-\epsilon Z^\prime_{\mu\nu} B^{\mu\nu}$ with $B^{\mu\nu}$ being the
field strength of the  $U(1)_Y$ gauge boson, is generally allowed by the gauge symmetry we consider. However, since the mixing does not generate
lepton-flavor-violating $b \to s \ell \ell$ and the parameter $\epsilon$ is constrained to be less than $10^{-2}$ for 
$m_{Z'}\sim 100$ GeV~\cite{Hook:2010tw}, we set $\epsilon=0$ for simplicity.

\section{NP contribution to $b \to s \mu\mu$ transition and constraints on the model}
\label{sec:NP}

\begin{figure}[tbp]
\begin{center}
\includegraphics[width=.85\textwidth]{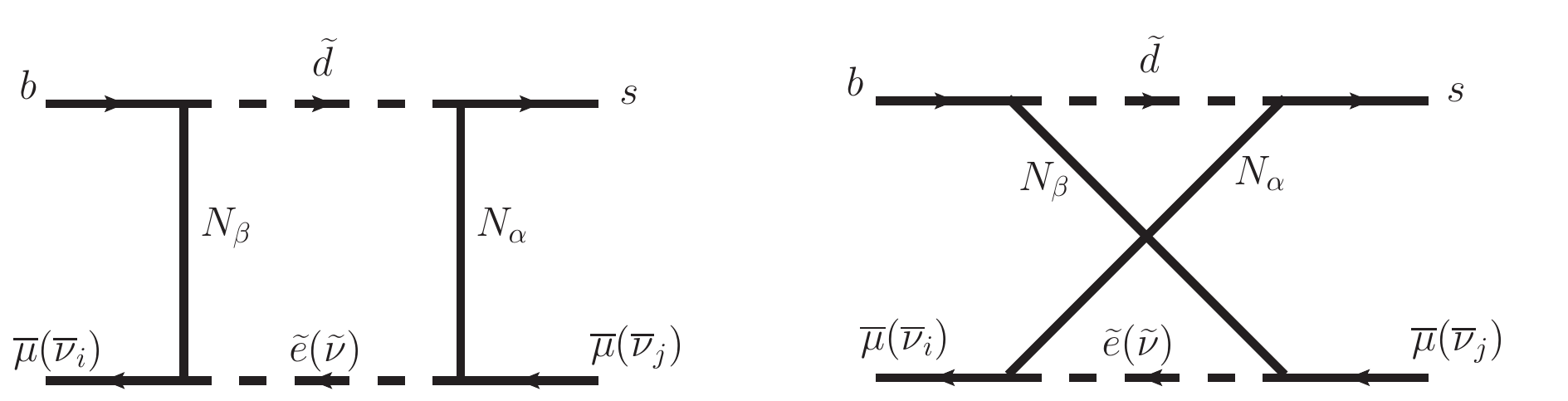}
\end{center}
\caption{Box diagrams generating $b \to s \mu^+ \mu^-$ and $b \to s \nu_k \overline{\nu}_l$ 
where $i,j(=1,2,3)$ are generation indices. In the figure $\alpha,\beta=\mp$.}
\label{fig:C9box}
\end{figure}

\begin{figure}[tbp]
\begin{center}
\includegraphics[width=.5\textwidth]{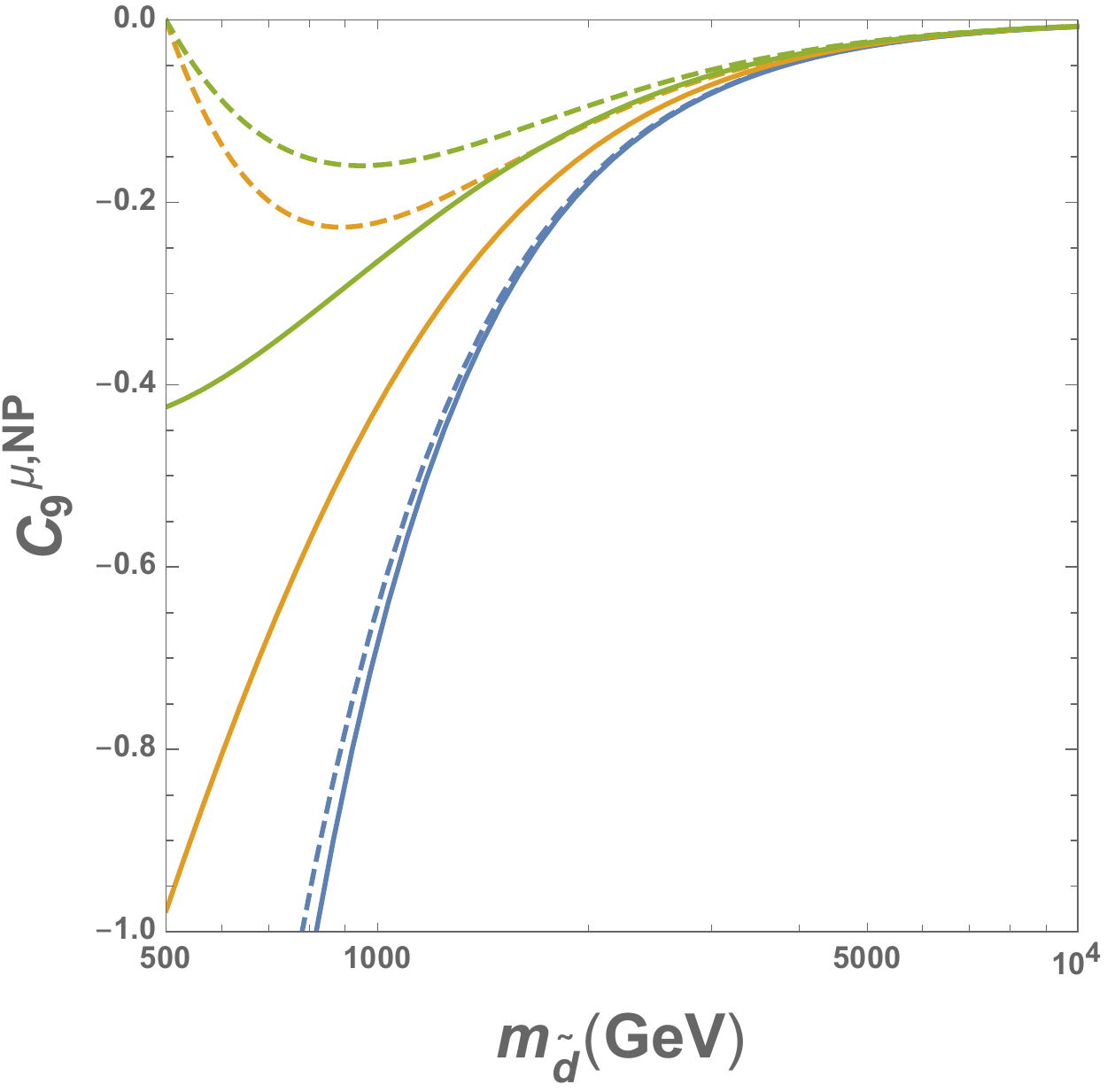}
\end{center}
\caption{Solid lines are $C_9^{\mu,{\rm NP}}$ as a function of $m_{\wt{d}}$ with three different choices, 
$(M_{-},M_{+})= (100,  100)$ GeV (blue), $(500,  500)$ GeV (orange), and $(500,1000)$ GeV  (green). For the plot we set
$m_{\wt{\mu}}=m_{\wt{d}}$. The corresponding dashed lines are obtained if we flipped the minus sign to plus in front of $j$ function
in (\ref{eq:C9}).}
\label{fig:C9_plot}
\end{figure}

In our model the $b \to s \mu^+ \mu^-$  transition operators $O_{9,10}^\mu$ which can explain the $b \to s \mu^+\mu^-$ anomaly
are generated via the box diagrams shown in figure~\ref{fig:C9box}. The arrows represent color or lepton number flow.
In the $U(1)_{L_\mu-L_\tau}$ model considered in~\cite{Baek:2017sew} the $b \to s \mu^+\mu^-$ transition operators are generated from
penguin diagrams at one-loop level. This clearly distinguishes the current model from the one considered in~\cite{Baek:2017sew}.
The existence of crossed diagrams in the right panel represents the Majorana nature of $N_\mp$.
The resulting $C_{9(10)}^{\mu,{\rm NP}}$, however, is not simple algebraic sum of the Majorana contributions presented in the literature,
{\it e.g.},  in~\cite{Arnan:2016cpy}.
In the limit $\Delta M \equiv M_+ - M_- \to 0$, the two Majorana fermions $N_\mp$ merges into a single Dirac fermion $N$. 
As a result, the crossed diagrams disappear in this limit due to the restored $U(1)_X$ symmetry.
This can be clearly seen from the minus sign in front of the second $j$ function in those Wilson coefficients
\begin{align}
C_9^{\mu,{\rm NP}} = -C_{10}^{\mu,{\rm NP}} &=-\frac{{\cal N} \lambda_q^2 \lambda_q^{3*}|\lambda_\ell^2|^2}{128 \pi \alpha_{\rm em} M_-^2} \Bigg[
 k(1,x_{\wt{d}-},x_{\wt{e} -}) +  k(x_{+-},x_{\wt{d}-},x_{\wt{e} -}) +x_{-+} k(1,x_{\wt{d}+},x_{\wt{e} +})  \nl
& +2  j(1,x_{\wt{d}-},x_{\wt{e}-})-4  j(x_{+-},x_{\wt{d}-},x_{\wt{e}-}) +2 x_{-+} j(1,x_{\wt{d}+},x_{\wt{e} +}) \Bigg],
\label{eq:C9}
\end{align}
where $x_{i\alpha} = m_i^2/M_\alpha^2 (i=\wt{d},\wt{\mu},\alpha=\mp)$ and ${\cal N}=\sqrt{2}/4 G_F V_{ts}^* V_{tb}$. 
Neglecting the minus sign which originates from the Majorana phase $\eta_-=-1$, the above results agree with those in \cite{Arnan:2016cpy}
up to overall sign. 
The loop functions $k$ and $j$ are listed in the appendix~\ref{app:loop}. In the limit of degenerate masses we get
$k(1,1,1)=1/3$ and $j(1,1,1)=-1/6$.
For $M_-=100$ GeV, $M_+=200$ GeV and $m_{\wt{d}}=m_{\wt{\ell}}=1$ TeV, we get
\begin{align}
C_9^{\mu,{\rm NP}} = -C_{10}^{\mu,{\rm NP}} &=-0.69 \left(\lambda_q^2 \lambda_q^{3*}  \over -0.15\right) \left(\lambda_\ell^2 \over
                                              2.4\right)^2,
\label{eq:C9_num}
\end{align}
which is close to the best fit value in (\ref{eq:C9_range}) to solve the $b \to s\mu\mu$ anomaly. 
To emphasise the importance of the {\it wrong} sign in (\ref{eq:C9}) and also to see the behaviour for large scalar quark/lepton masses, we show a
  plot of $C_9^{\mu,{\rm NP}}$, figure~\ref{fig:C9_plot}, as a function of $m_{\wt{d}}$ with three different choices, 
$(M_{-},M_{+})= (100,  100)$ GeV (solid blue), $(500,  500)$ GeV (solid orange), and $(500,1000)$ GeV  (solid green). For the plot we set
$m_{\wt{\mu}}=m_{\wt{d}}$. We can see $|C_9^{\mu,{\rm NP}}|$ is maximised when $M_{+}=M_{-}$.
The corresponding dashed lines are obtained if we flipped the minus sign to plus.
We can see the sign  flip drastically changes the result. We can also see the decoupling behaviour for heavy $m_{\wt{d}}$.

 The value $|\lambda_q^2 \lambda_q^{3*}|$
is constrained by the $B_s-\ol{B}_s$ mixing which will be considered below. 
As we will show later, a rather large value of $\lambda_\ell^2$ is
required to explain the anomaly. 
This will also affect the dark matter phenomenology as we will discuss later.

The $b \to s \mu\mu$ transition occurs also through $\gamma$- and $Z$-penguin diagrams. However, since $\gamma$ and
$Z$ couplings to leptons are flavour universal, the penguin contributions should not be too large.
We obtain
\begin{align}
C_9^{\ell,{\rm NP}} &=-\frac{{\cal N} e_d \lambda_q^2 \lambda_q^{3*}}{2 m_{\wt{d}}^2}
\Big[ P_\gamma(x_{-}) +P_\gamma(x_{+})\Big], \nl
C_{10}^{\ell,{\rm NP}} &=0,
\end{align}
where $e_d=-1/3$ and the loop-function $P_\gamma(x)$ is listed in the appendix~\ref{app:loop}.
We note that the above contribution to $C_9$ is independent of $\lambda_\ell^2$ and, compared to the box contribution, is suppressed in the case
when large $\lambda_\ell^2$ is required. For the benchmark point, $M_{-}=100$ GeV, $M_{+}=200$ GeV, and $m_{\wt{d}}=1$ TeV, we get
\begin{align}
C_9^{\ell,{\rm NP}}=2.0 \times 10^{-3} \left(\lambda_q^2 \lambda_q^{3*} \over -0.15\right),
\end{align}
which is 3 orders of magnitude smaller than $C_9^{\mu,{\rm NP}}$ in (\ref{eq:C9_num}). It is known that the $bsZ$-vertex is
proportional to $q^2$ with $q$ being the virtual 4-momentum of $Z$-boson~\cite{Arnan:2016cpy}. Since $q^2 \sim m_b^2$, the $Z$-penguin is further suppressed
compared with the $\gamma$-penguin by a factor $q^2/m_Z^2$.

Since $\wt{\ell}$ is $SU(2)_L$ doublet, the same box diagrams which generate $b \to s \mu \mu$ shown in figure~\ref{fig:C9box} also generate the
semi-leptonic decay $B \to K^{(*)} \nu \bar{\nu}$. The effective Hamiltonian is
\begin{align}
{\cal H}_{\rm eff}^{\nu_i \nu_j} &= -\frac{4 G_F}{\sqrt{2}} V_{ts}^* V_{tb} C_L^{ij} O_L^{ij},
\end{align}
where
\begin{align}
O_L^{ij} = \frac{e^2}{16 \pi^2} (\ol{s} \gamma^\mu P_L b) (\ol{\nu}_i \gamma_\mu (1-\gamma_5) \nu_j).
\end{align}
We can obtain $C_L^{22}$ just by replacing $m_{\wt{\ell}} \to m_{\wt{\nu}}$ and $\lambda_\ell^2 \to \lambda_\nu^2$. The effective
Yukawa coupling $\lambda_\nu^2$ is $\lambda_\nu^2 = U^*_{22} \lambda_\ell^2+ U^*_{32} \lambda_\ell^3$. Since neutrino flavors are not
measured in the experiments, the total branching ratio normalized to the SM
prediction defined by~\cite{Arnan:2016cpy}
\begin{align} 
R_{K^{(*)}}^{\nu\ol{\nu}} &= \frac{\sum_{i,j=1}^3\left|C_L^{\rm SM} \delta^{ij} + C_L^{ij}\right|^2}{3 \left|C_L^{\rm SM} \right|^2}, \quad \text{with} \; C_L^{\rm SM}
                            \approx -6.35,
\end{align} 
can be compared with the measurements
\begin{align}
%R_K^{\nu\ol{\nu}} < 4.3, \quad R_{K^*}^{\nu\ol{\nu}} < 4.4, \quad (\text{at 90\% C.L.}).
R_K^{\nu\ol{\nu}} < 4.8, \quad R_{K^*}^{\nu\ol{\nu}} < 6.2, \quad (\text{2$\sigma$}).
\end{align}
From the inequality,
\begin{align}
\left|1+\frac{C_L^{22}}{C_L^{\rm SM}}\right|^2 \le 3 R_{K}^{\nu\ol{\nu}} \le 14.4,  \;\; (\text{2$\sigma$}).
\end{align}
we get
\begin{align}
-17.7 \le C_L^{22} \le 30.4 \;\; (\text{2$\sigma$}).
\end{align}
This constraint is an order of magnitude weaker than the bound on the $SU(2)_L$-related $C_9^{\mu,{\rm NP}}$ given in (\ref{eq:C9_range}).

\begin{figure}[tbp]
\begin{center}
\includegraphics[width=.85\textwidth]{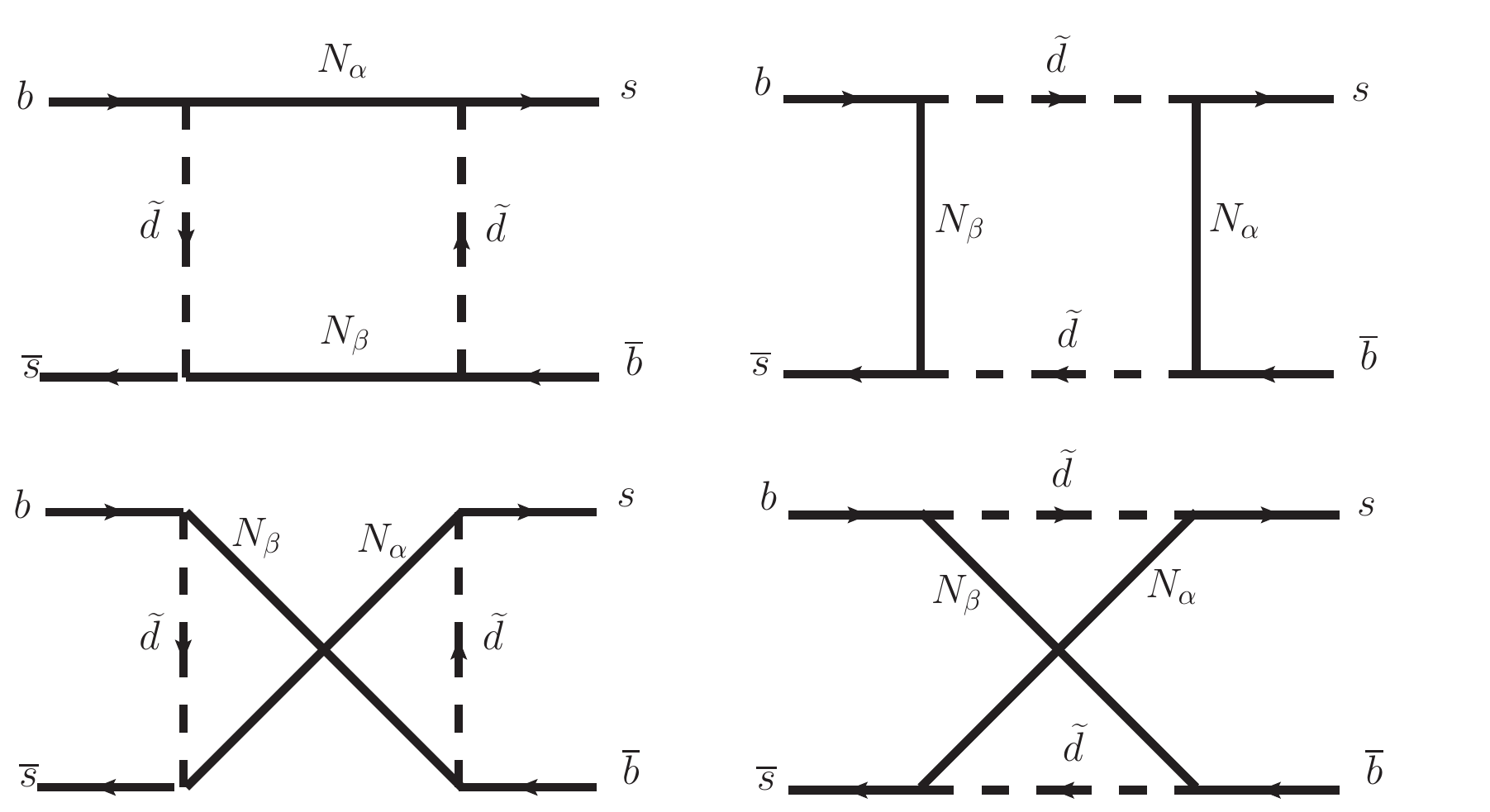}
\end{center}
\caption{Box diagrams generating $B_s-\ol{B}_s$ mixing. In the figure $\alpha,\beta=\mp$.}
\label{fig:BsBs}
\end{figure}

The $\wt{d}$ and $N_\mp$ also contribute to $B_s-\ol{B}_s$ mixing via the box
diagrams shown in figure~\ref{fig:BsBs}. The arrows stand for color flow. 
As in the case of $b \to s \mu^+ \mu^-$ diagrams in figure~\ref{fig:C9box}, the Majorana nature of dark fermions $N_\mp$ allows 
the crossed diagrams which vanish in the limit $M_+ \to M_-$.
The effective Hamiltonian for the $B_s-\ol{B}_s$ mixing 
\begin{align}
{\cal H}_{\rm eff}^{\Delta B=2} &= C_1    (\ol{s} \gamma_\mu P_L b) (\ol{s}  \gamma^\mu P_L b), 
\end{align}
has both the SM contribution and the NP contributions
\begin{equation}
C_1 = C_1^{\rm SM} + C_1^{\rm NP}.
\end{equation}
The SM Wilson coefficient at the electroweak scale is
\begin{align}
C_1^{\rm SM} &= \frac{G_F m_W^2}{4 \pi^2} (V_{ts}^* V_{tb})^2 S_0(x_t), 
\label{eq:BsBsSM}
\end{align}
where $x_t = m_t^2/m_W^2$ and the loop function $S_0(x_t)$ can be found, {\it e.g.}, in~\cite{Buchalla:1995vs}.
The NP contribution from figure~\ref{fig:BsBs}  reads
\begin{align}
C_1^{\rm NP} &= \frac{(\lambda_q^2 \lambda_q^{3*})^2}{512 \pi^2 m^2_{\wt{d}}} \sum_{\alpha,\beta=\mp}\Big[
k(1,y_\alpha,y_\beta)+2 \eta_\alpha \eta_\beta \sqrt{y_\alpha y_\beta} j(1,y_\alpha,y_\beta)
\Big],
\end{align}
where $y_\alpha=M_\alpha^2/m^2_{\wt{d}}, \, (\alpha=\mp)$ and $\eta_\mp=\mp 1$ are Majorana phases.
We have checked the above expression agrees with the corresponding one in~\cite{Arnan:2016cpy} if we set
$M_+=M_-$ and $\eta_-=\eta_+=1$.
We note that the crossed diagrams in figure~\ref{fig:BsBs} disappear in the limit $M_+ \to M_-$ because $N_\mp$ merges into a single Dirac fermion
as in the case of figure~\ref{fig:C9box}. And we should not use  the results in \cite{Arnan:2016cpy} naively.
The allowed range for $C_1^{\rm NP}$ from the measurement of the mass difference in the $B_s-\ol{B}_s$ system is~\cite{Arnan:2016cpy} 
\begin{align}
-2.1 \times 10^{-11} \le C_1^{\rm NP} \le 0.6 \times 10^{-11} \, {\rm GeV}^{-2},
%C_1^{\rm NP}&= (-2.8, 1.3) \times 10^{-5} \, {\rm TeV}^{-2} \; (3\sigma).
\label{eq:BsBs}
\end{align}
at 2$\sigma$ level. For real $\lambda_q^2 \lambda_q^{3*}$ and $M_-\le m_{\wt{d}}$, $C_1^{\rm NP}$ is always positive. 
For example, for $M_-=100$ GeV, $M_+=200$ GeV, $m_{\wt{d}}=1.67$ TeV,  and $\lambda_q^2 \lambda_q^{3*} =-0.15$, the upper bound is
saturated:
\begin{align}
C_1^{\rm NP} = 0.60 \times 10^{-11} \left( \lambda_q^2 \lambda_q^{3*} \over -0.15 \right)^2 \; {\rm GeV}^{-2}.
\label{eq:lambda_BsBs}
\end{align}

\begin{figure}[tbp]
\begin{center}
\includegraphics[width=.5\textwidth]{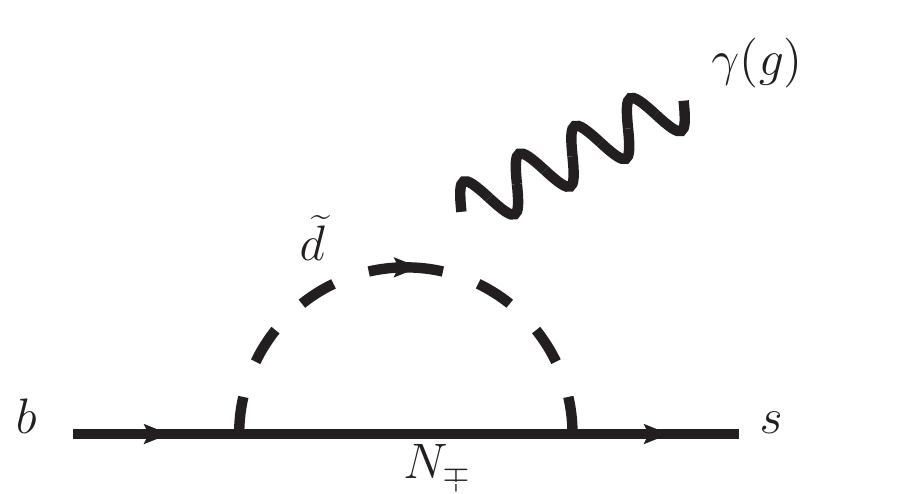}
\end{center}
\caption{Feynman diagrams generating $b \to s \gamma  (g)$. The photon (gluon) line, $\gamma (g)$, can be attached to any
  charged (colored) particles.}
\label{fig:bsr}
\end{figure}

The Yukawa interactions with couplings $\lambda_q^2$ and $\lambda_q^3$ also generate operators $O_{7\gamma}$ and $O_{8g}$, contributing 
to a radiative flavor-changing $b$ decay, $b \to s \gamma$. 
The experimental measurement and the SM prediction of the inclusive branching fraction of radiative $B$-decay, $\ol{B} \to X_s \gamma$,
are~\cite{Amhis:2016xyh, Misiak:2015xwa} 
\begin{align}
{\cal B}\left[\ol{B} \to X_s \gamma, \left(E_\gamma > 1.6 \,{\rm GeV}\right) \right]^{\rm exp} &= (3.32 \pm 0.16) \times 10^{-4}, \nl
{\cal B}\left[\ol{B} \to X_s \gamma, \left(E_\gamma > 1.6 \,{\rm GeV}\right)\right]^{\rm SM} &= (3.36 \pm 0.23)  \times 10^{-4} .
\label{eq:bsr}
\end{align}
The NP contribution to $C_7^\gamma$ at the electroweak scale whose diagram is shown in figure~\ref{fig:bsr} is obtained to be
\begin{align}
C_{7\gamma}^{\rm NP}&= \frac{{\cal N} \lambda_q^2 \lambda_q^{3*}}{4 m^2_{\wt{d}}} e_d [J_1(y_-)+J_1(y_+)], \nl
C_{8g}^{\rm NP}&= \frac{{\cal N} \lambda_q^2 \lambda_q^{3*}}{4 m^2_{\wt{d}}}  [J_1(y_-)+J_1(y_+)], 
\end{align}
where $e_d=-1/3$ is the electric charge of $\wt{d}$, $y_\mp =M_\mp^2/m^2_{\wt{d}}$, and the loop function $J_1(y)$ is given in the
appendix~\ref{app:loop}.
The ratio~\cite{Misiak:2015xwa} 
\begin{align}
R_{b \to s\gamma} = \frac{{\cal B}^{\rm exp}(b \to s\gamma)}{{\cal B}^{\rm SM}(b \to s\gamma)}-1 
= -2.45 \left( C_{7\gamma}^{\rm NP}+0.24 C_{8g}^{\rm NP} \right),
\end{align}
which includes the QCD effect can constrain the combination 
$ C_{7\gamma}^{\rm NP}+0.24 C_{8g}^{\rm NP}$. Using (\ref{eq:bsr}), we obtain
\begin{align}
-0.065 \le C_{7\gamma}^{\rm NP}+0.24 C_{8g}^{\rm NP} \le 0.073,
\label{eq:bsr_constraint}
\end{align}
at 2$\sigma$ level. For $M_-=100$ GeV, $M_+=200$ GeV, $m_{\wt{d}}=1$ TeV, we obtain
\begin{align}
C_{7\gamma}+0.24 C_{8g} =-4.1 \times 10^{-4} \left(\lambda_q^2 \lambda_q^{3*} \over -0.15\right),
\end{align}
which is at least two orders of magnitude below the current bounds in (\ref{eq:bsr_constraint}).

The leptonic decay $B_s \to \mu^+\mu^-$ can be very sensitive to NP models such as 
Minimal Supersymmetric Standard Model (MSSM)~\cite{Babu:1999hn,Baek:2002rt,Baek:2002wm,Baek:2004tm,Baek:2004et}.
In our model the contributions to the scalar operators, $(\ol{s}_L b_R)(\ol{\ell}\ell)$, 
$(\ol{s}_L b_R)(\ol{\ell} \gamma_5\ell)$, are suppressed by small muon mass compared to NP scale. 
However, since the model explains $b \to s\mu\mu$ anomaly in the $C_9 = -C_{10}$ scenario,
NP contribution to $C_{10}$ can be sizeable. The measurement by LHCb collaboration~\cite{Aaij:2017vad}
\begin{align}
{\cal B}(B_s \to \mu^+ \mu^-)^{\rm LHCb} = (3.0 \pm 0.6^{+0.3}_{-0.2}) \times 10^{-9},
\label{eq:Bsmm_exp}
\end{align}
is a little smaller than the SM prediction~\cite{Bobeth:2013uxa,Bobeth:2013tba}
\begin{align}
{\cal B}(B_s \to \mu^+ \mu^-)^{\rm SM}= (3.65 \pm 0.23) \times 10^{-9},
\label{eq:Bsmm_SM}
\end{align}
although they agree with each other within 1$\sigma$.
Interestingly the scenario in (\ref{eq:C9_range}) requires positive $C_{10}^{\mu,{\rm NP}}$ in the 2$\sigma$ range whereas $C_{10}^{\rm SM}$
is negative, predicting smaller branching fraction than that of the SM as favoured by the LHCb experiment.
Using (\ref{eq:Bsmm_exp}) and (\ref{eq:Bsmm_SM}), we obtain
the ratio,
\begin{align}
\frac{{\cal B}(B_s \to \mu^+\mu^-)^{\rm LHCb}}{{\cal B}(B_s \to \mu^+ \mu^-)^{\rm SM}}=
\frac{|C_{10}^{\rm SM}+C_{10}^{\mu,{\rm NP}}|^2}{|C_{10}^{\rm SM}|^2}=0.82 \pm 0.19.
\end{align}
From $C_{10}^{\rm SM}=-4.1$ we can read the 2$\sigma$ range for $C_{10}^{\mu,{\rm NP}}$,
\begin{align}
-0.45 \le C_{10}^{\mu,{\rm NP}} \le 1.2, 
\end{align}
in which the whole 2$\sigma$ range for $C_{10}^{\mu,{\rm NP}}$ in (\ref{eq:C9_range}) is contained.
This shows the region which solves $b \to s \mu\mu$ anomaly automatically satisfies the ${\cal B}(B_s \to \mu^+\mu^-)$ constraint.

The anomalous magnetic moment of muon $a_\mu=(g-2)_\mu/2$ also receives contribution from the diagram figure~\ref{fig:bsr} with replacement
$b,s \to \mu$ and $\wt{d} \to \wt{e}$. The effective Hamiltonian for $a_\mu$ is
\begin{align}
{\cal H}^{a_\mu}_{\rm eff} &=-\frac{a_\mu e}{4 m_\mu} (\ol{\mu} \sigma^{\mu\nu} \mu) F_{\mu\nu}.
\label{eq:amu_H}
\end{align}
We get
\begin{align}
a_\mu^{\rm NP} &= -\frac{|\lambda_\ell^2|^2 m_\mu^2}{(4\pi)^2 m^2_{\wt{e}}} \left(J_1(y_-)+J_1(y_+) \right),
\label{eq:amu_NP}
\end{align}
where $y_\mp=M_\mp^2/m^2_{\wt{e}}$ and the loop function $J_1(y)$ is listed in appendix~\ref{app:loop}. The difference between the
experimental measurement~\cite{Bennett:2006fi} and a most recent SM prediction~\cite{Kurz:2014wya}
\begin{align}
\Delta a_\mu&=a_\mu^{\rm exp}-a_\mu^{\rm SM} = (236 \pm 87) \times 10^{-11},
\label{eq:amu_diff}
\end{align}
shows $2.7\sigma$ discrepancy. The result (\ref{eq:amu_NP}) is opposite in sign to the deviation in (\ref{eq:amu_diff}).
We can use (\ref{eq:amu_diff})  as a constraint. 
The NP contribution to $a_\mu$ for the benchmark point $M_-=100$ GeV, $M_+=200$ GeV, $m_{\wt{e}}=1$ TeV, 
\begin{align}
a_\mu^{\rm NP}=-6.5 \times 10^{-11} \left(\lambda^2_\ell \over 2.4\right)^2,
\end{align}
can satisfy the 3$\sigma$ range in the discrepancy:   $-25 \times 10^{-11} \le \Delta a_\mu \le 497 \times 10^{-11}$.
Note that in our model a suppression factor $m_\mu^2/m_{\wt{e}}^2$  results from the chirality flip in the external muon line.

Additional constraint may come from effective $Z\mu^+\mu^-$ vertex which is generated from the diagrams similar to figure~\ref{fig:bsr} but
with the replacement: $\gamma \to Z$, $\wt{d} \to \wt{e}$ and $b,s \to \mu$. The NP contribution $g_L^{\rm NP}$ to the effective 
vertex~\cite{Baek:2002xf} given by
\begin{align}
{\cal L}_{\rm eff}=-\frac{g}{\cos\theta_W} (g_{L_\mu}^{\rm SM}+g_{L_\mu}^{\rm NP}) Z_\alpha \ol{\mu_L} \gamma^\alpha \mu_L,
\end{align}
turns out to be finite and proportional to the four-momentum square of $Z$, $q^2$:
\begin{align} 
\frac{g_{L_\mu}^{\rm NP}}{g_{L_\mu}^{\rm SM}}(q^2)=-\frac{|\lambda_\ell^2|^2 q^2}{32 \pi^2 m^2_{\wt{e}}} \left(\wt{F}_9(y_-) +\wt{F}_9(y_+)\right),
\end{align} 
where $y_\mp=M_\mp^2/m^2_{\wt{e}}$. The loop function $\wt{F}_9(y)$ can be found in the appendix~\ref{app:loop}. 
The SM contribution at tree level is $g_{L_\mu}^{\rm SM, tree}=T_3^{\mu}-Q^\mu\sin^2\theta_W=-1/2+\sin^2\theta_W$.
The LEP experiment measured the coupling $g_{L_\mu}$ at $Z$-pole~\cite{ALEPH:2005ab} with the result
\begin{align}
g_{L_\mu}^{\rm exp} = -0.2689 \pm 0.0011.
\end{align}
We impose the constraint 
\begin{align}
\left| \frac{g_{L_\mu}^{\rm NP}}{g_{L_\mu}^{\rm SM}}(M_Z^2)\right| \le 0.8 \%, 
\label{eq:gL_exp}
\end{align}
at 2$\sigma$ level. For $M_-=100$ GeV, $M_+=200$ GeV, $m_{\wt{e}}=1$ TeV, we obtain
\begin{align}
\frac{g_{L_\mu}^{\rm NP}}{g_{L_\mu}^{\rm SM}}(M_Z^2) =0.085 \left(\lambda_\ell^2 \over 2.4\right)^2  \%,
\end{align}
which is an order of magnitude below the experimental upper bound (\ref{eq:gL_exp}).
As we have seen above, the constraint from the $B_s-\ol{B}_s$ mixing is the strongest and all the others are orders of magnitude below the
current experimental bound. We'll impose only the $B_s-\ol{B}_s$ mixing constraint for our numerical analysis.

The model is also constrained by collider experiments such as the LHC. The most telltale signature for the model is the observation
of new scalar particles.
For example, the new colored-scalar particles can be searched for at the LHC via, $p p \to \wt{d} \wt{d}^*$ followed by the
decay $\wt{d} \to b N_-$  processes, giving $b$-jets and missing transverse momentum events~\cite{Baek:2016lnv,Baek:2017ykw}. 
Using 36.1 fb$^{-1}$  of $ p p $ collision data at $\sqrt{s} = 13$ TeV the ATLAS collaboration excludes 
$m_{\wt{d}} \lesssim 950$ GeV for $M_- \lesssim 420$ GeV at 95\% confidence level~\cite{Aaboud:2017wqg}.
To be conservative we use $m_{\wt{q}(\wt{\ell})}  \ge 1$ TeV in the numerical analysis below.

\section{Dark matter phenomenology and numerical results}
\label{sec:DM}

\begin{figure}[tbp]
\begin{center}
\includegraphics[width=.85\textwidth]{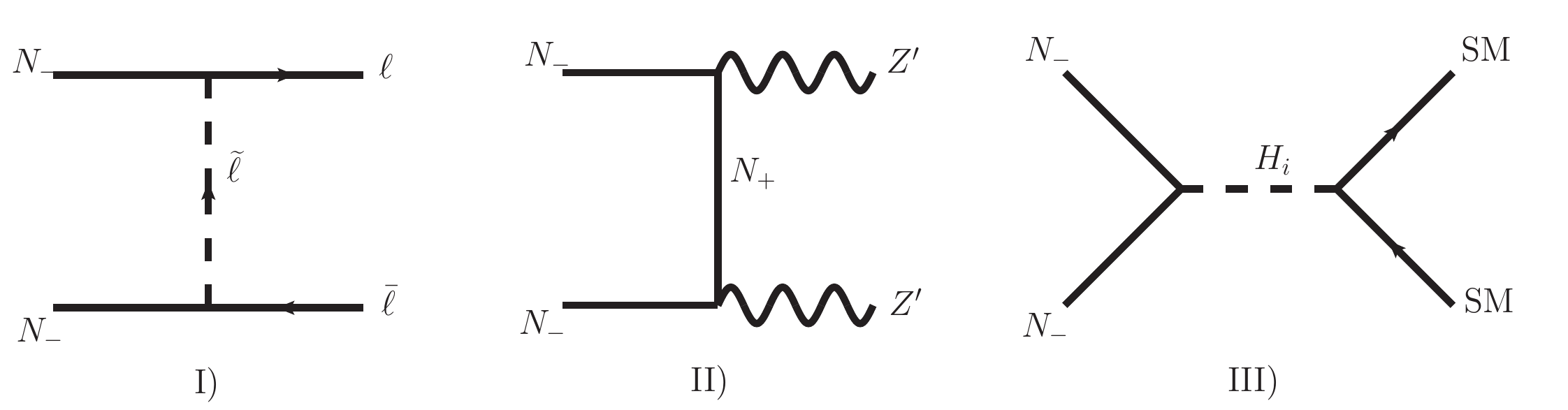}
\end{center}
\caption{Representative diagrams for DM annihilation in scenarios I), II), and III).}
\label{fig:DM_ann}
\end{figure}

In this section we discuss dark matter physics such as dark matter relic abundance and direct detection for our dark matter candidate $N_-$. 
In our model weakly interacting massive particle  (WIMP) $N_{-}$ is a good candidate for a thermal dark matter. We assume
  $N_{-}$ is the only dark matter component.  
In the early universe the dark matter $N_{-}$ is in thermal equilibrium with the SM plasma through processes, 
$N_- \, N_- \leftrightarrow {\rm  SM} \, {\rm SM}$, some of which are shown in figure~\ref{fig:DM_ann}. 
As the universe cools down and the rates for these processes drop below the expansion rate, the dark matter particle freezes out from the
thermal bath. The current relic density can be estimated from
\begin{align}
\Omega_{\rm DM} h^2 \approx \frac{3 \times 10^{-27} \,{\rm cm^3/s}}{\langle \sigma v \rangle}.
\end{align}
We will see that there is strong interplay between $b \to s \mu\mu$ anomaly and dark matter phenomenology. 
The dominant DM interactions
\begin{align}
&\Delta {\cal L} \nl
=&-\frac{1}{\sqrt{2}} \lambda_\ell^2 \ol{\ell}^2_L \wt{\ell} (N_-+N_+) +h.c.\nl
-& g_X Q Z'_\mu \ol{N}_+ \gamma^\mu N_- \nl
-&\frac{f}{2\sqrt{2}} (-H_1 \sin\alpha_H + H_2 \cos\alpha_H) ( -\ol{N}_- N_- +\ol{N}_+ N_+)  
-(H_1 \cos\alpha_H + H_2 \sin\alpha_H) \sum_f \frac{m_f}{v_H} \ol{f} f,
\end{align}
include Yukawa interaction with the 2nd generation leptons, dark-gauge interaction, and Higgs portal interaction.
We study three benchmark scenarios depending on the dominant interactions contributing to the dark matter annihilation: 
I) dark-Yukawa-interaction($\lambda_\ell^2$)-dominated channels, II)
dark-gauge-interaction($g_X$)-dominated channels and III) Higgs-portal($\lambda_{HS}$)-dominated channels. Representative diagrams for each
category are shown in figure~\ref{fig:DM_ann}.

\begin{figure}[tbp]
\begin{center}
\includegraphics[width=.45\textwidth]{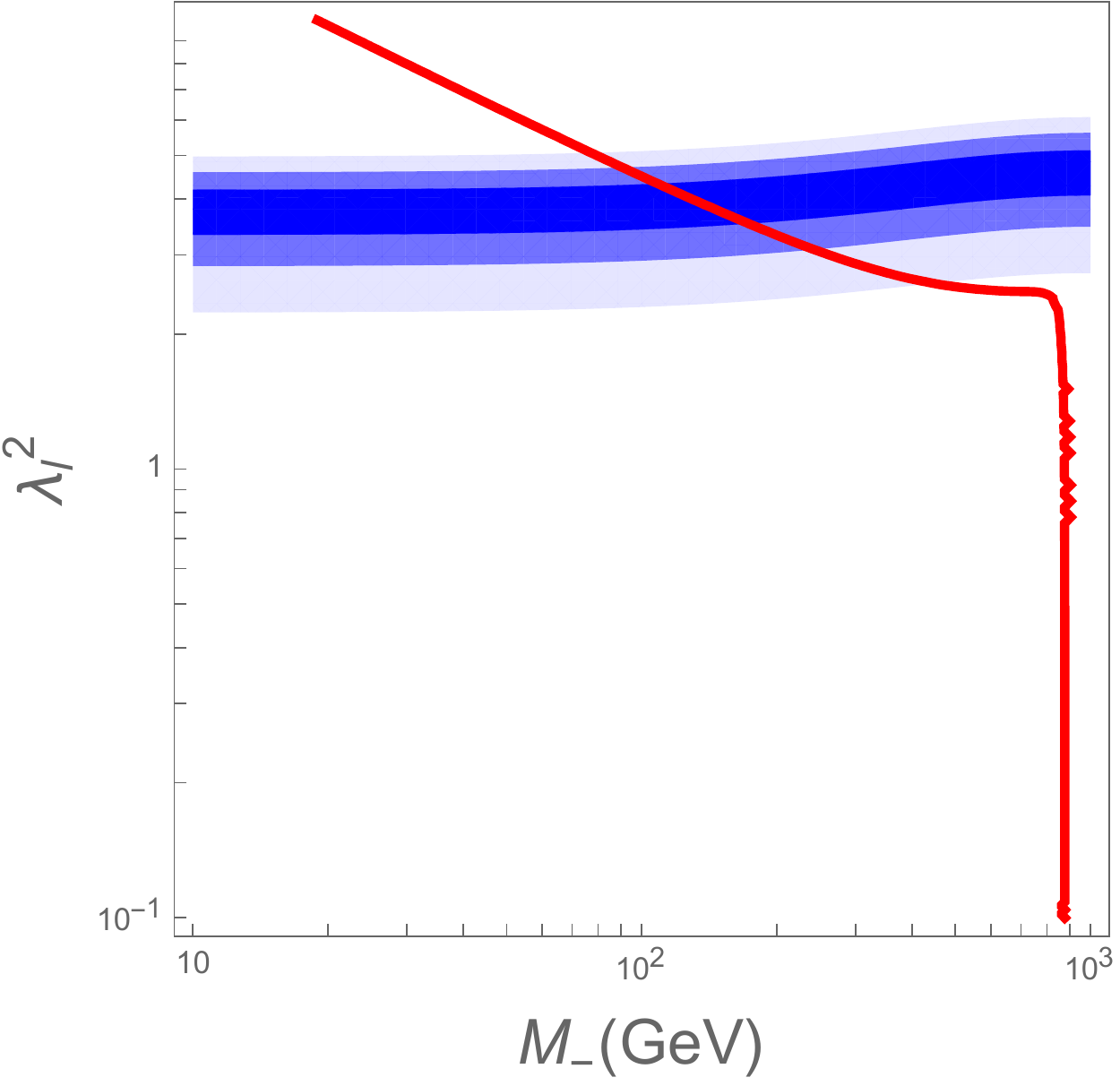}
\includegraphics[width=.45\textwidth]{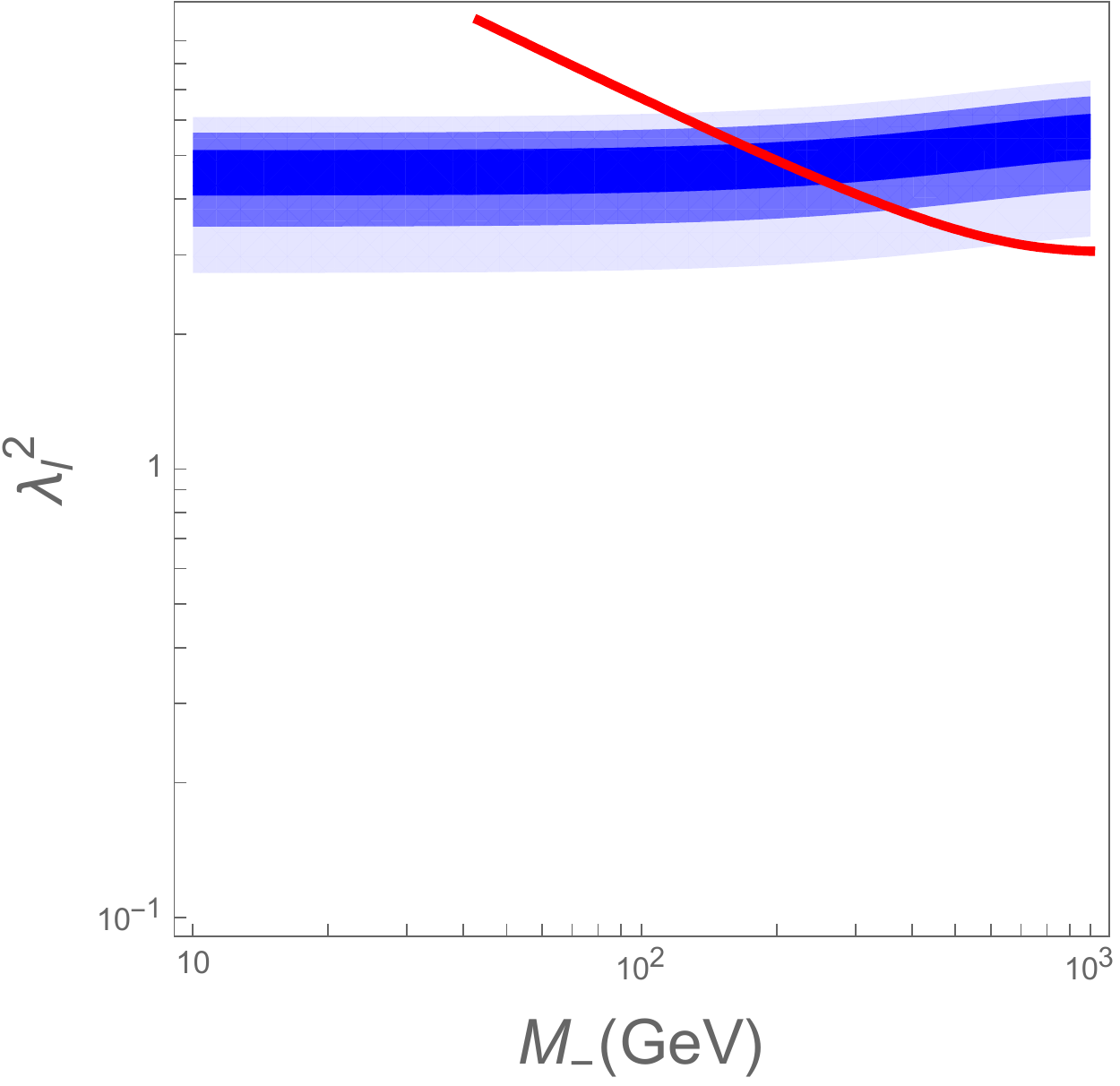}
\end{center}
\caption{DM relic density and $C_9^{\mu,{\rm NP}}$ in  $(M_-,\lambda_{\ell}^2)$-plane in Scenario I).   
The red line is a contour line for the constant dark matter relic density $\Omega_{\rm DM} h^2 = 0.12$.
The dark (medium, light) blue region satisfies $1 \sigma (2\sigma, 3\sigma)$  allowed region for $C_9^{\mu,{\rm NP}}$ 
to explain $b \to s \mu\mu$ anomaly in (\ref{eq:C9_range}).
For the left (right) plot, we take $\lambda_{q}^2=-0.2 (-0.3)$, $M_+=m_{\wt{q}}=m_{\wt{\ell}} =1000 (1500)$ GeV. 
The other parameters we take can be found in the text.
}
\label{fig:Yul}
\end{figure}

In the Scenario I) the diagram of type I) in figure~\ref{fig:DM_ann} can easily dominate over other channels because we need relatively large
$\lambda_{\ell}^2$ to explain the $b \to s \mu\mu$ anomaly since $\lambda_{q}^2 \lambda_{q}^{3*}$ is strongly constrained by
the mass difference in $B_s-\ol{B}_s$ mixing as shown in (\ref{eq:lambda_BsBs}).
figure~\ref{fig:Yul} shows DM relic density and $C_9^{\mu,{\rm NP}}$ in  $(M_-,\lambda_{\ell}^2)$-plane in Scenario I).   
The red line is a contour line for a constant dark matter relic density $\Omega_{\rm DM} h^2 = 0.12$~\cite{Ade:2015xua}.
The dark (medium, light) blue region satisfies $1 \sigma (2\sigma, 3\sigma)$  allowed region for $C_9^{\mu,{\rm NP}}$ 
to explain $b \to s \mu\mu$ anomaly in (\ref{eq:C9_range}).
For the left (right) plot, we take $\lambda_{q}^2=-0.2 (-0.3)$, $M_+=m_{\wt{q}}=m_{\wt{\ell}} =1000 (1500)$ GeV. 
For figure~\ref{fig:Yul} we fixed the other parameters as
\begin{align}
&g_X=\alpha_H =0.01, \quad \lambda_{q}^3 = 0.5, \quad m_{Z'} =m_{H_2} =2000 \, {\rm GeV}.
\end{align}
We take  the rest free parameters as
\begin{align}
&Q =1, \quad \lambda_{\wt{q}} =\lambda_{\wt{\ell}} =\lambda_{H\wt{q}} =\lambda_{H\wt{\ell}} =\lambda_{S\wt{q}}=\lambda_{S\wt{\ell}}=0.5, \nl
&\lambda^\prime_{H\wt{q}} =\lambda^{\prime\prime}_{H\wt{q}} =\lambda^\prime_{H\wt{\ell}} =\lambda^{\prime\prime}_{H\wt{\ell}} =0.
\label{eq:scan_fixed}
\end{align}
for all scenarios.
We note that the direct detection constraint on Scenarios I) is not significant because it first occurs at one-loop process
and is also proportional to the momentum transfer to nucleon.
To be more specific, let us consider $N_{-}q \to N_{-}q$ process mediated by the dominant one-loop $N_{-}N_{-}\gamma$ vertex inside which
$\mu$ and $\wt{e}$ are running. A naive dimensional analysis gives the effective operator
\begin{align}
{\cal L}_{eff} \sim \frac{\alpha_{\rm em} |\lambda_\ell^2|^2}{4 \pi m_{\wt{e}}^2} (\ol{N}_{-} \gamma^\mu \gamma_5 N_{-})(\ol{q} \gamma_\mu
  q),
\label{eq:DD1}
\end{align}
where we used the Majorana nature of $N_-$.
In the zero momentum transfer limit only the space component of $\ol{N}_{-} \gamma^\mu \gamma_5 N_{-}$ and the time component of  
$\ol{q} \gamma_\mu q$ survive. Consequently for non-zero but small momentum transfer, 
the contribution of the above operator to the direct detection cross section is suppressed 
by $Q^2/M_{-}^2$, where $Q \approx m_N v \approx 1$ MeV is the maximum momentum transfer to a nucleon~\cite{Belanger:2008sj}.
Since this suppression factor $Q^2/M_{-}^2 \sim 10^{-10}$ (for $M_{-} \sim 100$ GeV) is very small, we can safely neglect the contribution
to the direct detection.
Even if we set $\lambda_{q}^1 \equiv 0$, $\lambda_u^1$ is induced as we saw in (\ref{eq:lambda_u1}). We may expect the effective
$uuZ'$ vertex is generated by $\lambda_u^1$ at one-loop level. It has contribution to the direct detection cross section via
the operator
\begin{align}
{\cal L}_{eff} \sim \frac{\alpha_{X} |\lambda_u^1|^2}{4 \pi m_{Z'}^2} (\ol{N}_{+} \gamma^\mu  N_{-})(\ol{q} \gamma_\mu q),
\label{eq:DD2}
\end{align}
which is loop- and CKM-suppressed. 
Most importantly the DM scattering off the nucleon in this case is inelastic upward scattering which does not occur
unless the mass splitting is less than 1 keV: $\Delta M=M_+-M_- \lesssim m_N v^2 \sim 1$ keV.

We have also checked that the entire region satisfies bound from the $B_s-\ol{B}_s$ mixing which is the strongest flavor 
constraint.
Thus we can explain the $b \to s \mu\mu$ anomaly and the correct DM relic abundance of the universe while
satisfying the constraints from the flavor physics, astrophysics, and cosmology.
In both plots of figure~\ref{fig:Yul} we can see that the $N_- N_- \to \ell \ol{\ell}$ process in figure~\ref{fig:DM_ann} I) determines the
DM relic abundance in almost all the mass range considered. 
In the left plot  the $\Omega_{\rm DM} h^2$ contour line  near $M_- \approx M_+ =1000$ GeV drops abruptly because
the coannihilation processes, such as $N_- N_+ \to Z' \to \wt{\ell} \wt{\ell}^*$, which do not depend on $\lambda_{\ell}^2$ can dominate.
In the right plot these coannihilation processes cannot occur due to large mass difference $\Delta M$.
The annihilation cross section of $N_- N_- \to \ell \ol{\ell}$ is $p$-wave suppressed and is approximately given by~\cite{Baek:2015fma}
\begin{align}
\sigma v (N_- N_- \to \ell \ol{\ell}) \simeq \frac{\lambda_{\ell}^2 M_-^2 (M_-^4+m_{\wt{\ell}}^4) v^2}{96 \pi (M_-^2+m_{\wt{\ell}}^2)^4}
  + O(v^4),
\label{eq:NN2ll}
\end{align}
where we set $m_{\wt{\ell}} \equiv m_{\wt{\nu}}=m_{\wt{e}}$.
For fixed $m_{\wt{\ell}}$ and $\lambda_{\ell}^2$ the above annihilation cross section increases as $M_-$ increases, which is the reason the 
$\lambda_\ell^2$ is decreasing as $M_-$ is increasing along the red lines in figure~\ref{fig:Yul}. 

\begin{figure}[th]
\begin{center}
\includegraphics[width=.45\textwidth]{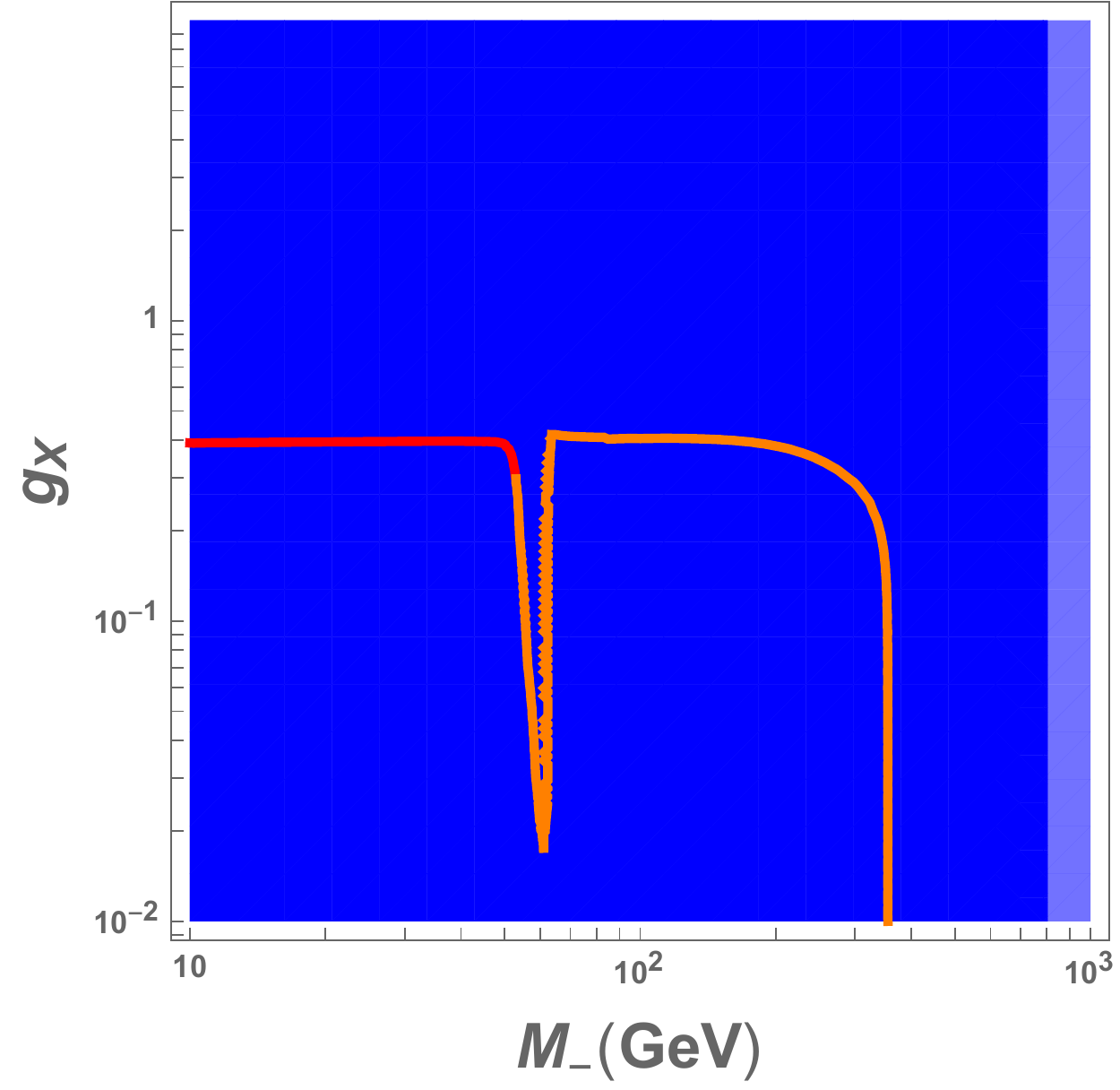}
\includegraphics[width=.45\textwidth]{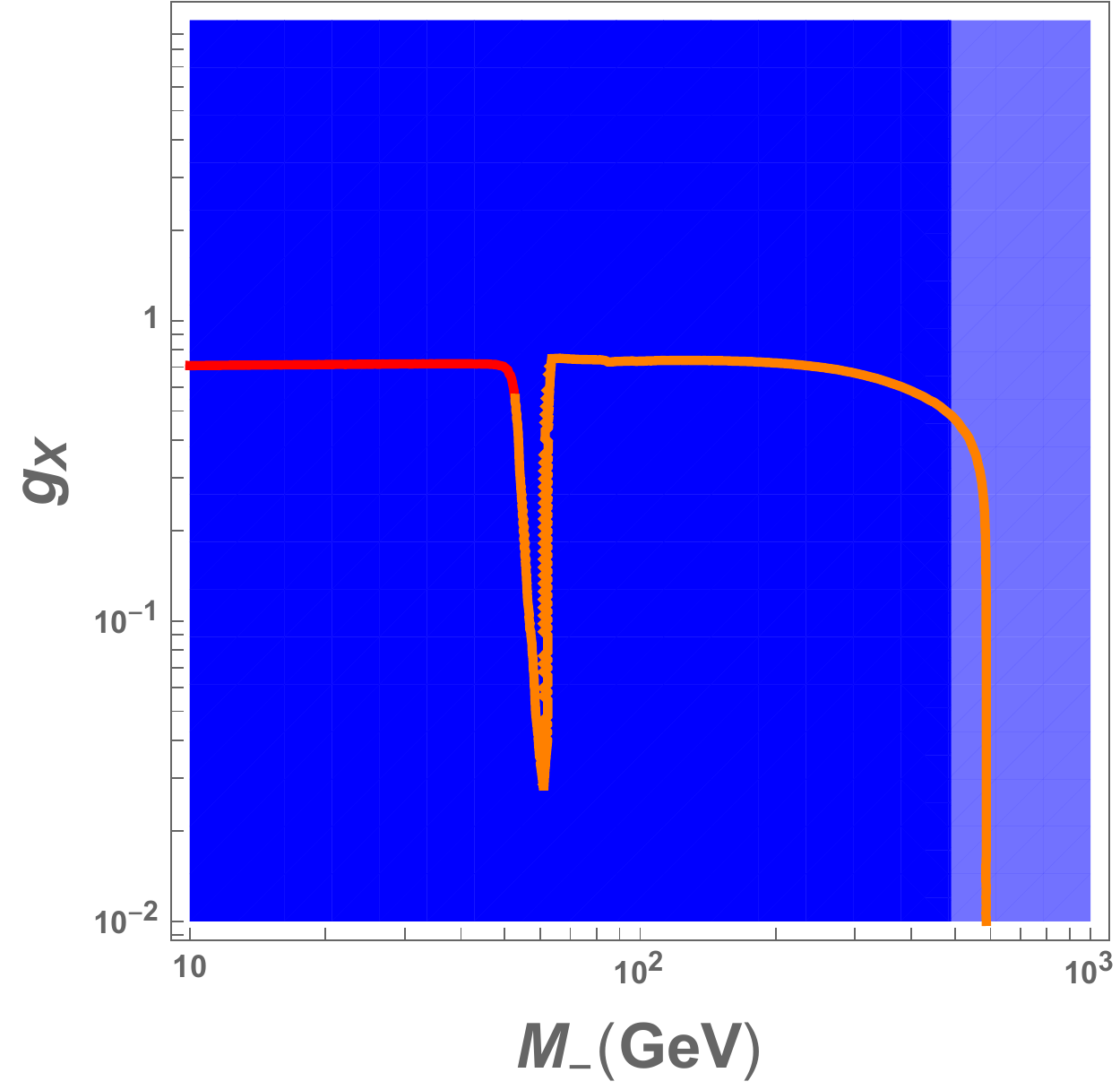}
\end{center}
\caption{DM relic density and $C_9^{\mu,{\rm NP}}$ in  $(M_-,g_X)$-plane in Scenario II).   
The red-orange line is a contour line for the constant dark matter relic density $\Omega_{\rm DM} h^2 = 0.12$.
The $N_-N_- \to Z'Z'$ dominates in the red segment where $M_- \lesssim 50$ GeV, 
while $N_- N_- \to H_1 \to {\rm SM} {\;\rm SM}$ (near resonance region) or $N_- N_- \to \mu^- \mu^+ (\nu_\mu \ol{\nu}_\mu)$ 
($M_- \gtrsim 65$ GeV) becomes important in  the orange segment.
Therefore only the line segment with red colour represents Scenario II).
The dark (medium) blue region satisfies $1 \sigma (2\sigma)$  allowed region for $C_9^{\mu,{\rm NP}}$ 
to explain $b \to s \mu\mu$ anomaly in (\ref{eq:C9_range}).
For the left (right) plot, we take $\lambda_{q}^2=-0.3 (-0.4)$, $m_{\wt{q}}=m_{\wt{\ell}} =2000 (2500)$ GeV. 
The other parameters we take can be found in the text.
}
\label{fig:Zp}
\end{figure}
In scenario II), the $Z'$ is lighter than the dark matter and the scalar-quark and scalar-lepton
are very heavy so that its contribution to the DM annihilation is suppressed. 
Thus we expect that the type of diagrams in figure~\ref{fig:DM_ann} II)
dominates the DM annihilation. 
Figure~\ref{fig:Zp} shows plots for DM relic density and $C_9^{\mu,{\rm NP}}$ in  $(M_-,g_X)$-plane in Scenario II).   
For the left (right) plot, we take $\lambda_{q}^2=-0.3 (-0.4)$, $M_+=1000, m_{\wt{q}}=m_{\wt{\ell}} =2000 (2500)$ GeV. 
For the other parameters we take 
\begin{align}
&\alpha_H =0.01, \quad \lambda_{q}^3 = 0.5, \quad \lambda_\ell^2=5.0, \quad
M_+=1000 \, {\rm GeV}, \quad m_{H_2} =2000 \,{\rm GeV},
\end{align}
and (\ref{eq:scan_fixed}). We set $m_{Z'} \equiv M_-/2$ so that $N_- N_- \to  Z' Z'$ is kinematically allowed.
The red-orange line is a contour line for a constant dark matter relic density $\Omega_{\rm DM} h^2 = 0.12$.
The $N_-N_- \to Z'Z'$ dominates in the red segment where $M_- \lesssim 50$ GeV, 
while $N_- N_- \to H_1 \to {\rm SM} {\;\rm SM}$ (near resonance region) or $N_- N_- \to \mu^- \mu^+ (\nu_\mu \ol{\nu}_\mu)$ 
($M_- \gtrsim 65$ GeV) becomes important in  the orange segment.
Therefore only the line segment with red colour represents Scenario II).
The dark (medium) blue region satisfies $1 \sigma (2\sigma)$  allowed region for $C_9^{\mu,{\rm NP}}$ 
to explain $b \to s \mu\mu$ anomaly in (\ref{eq:C9_range}).
We note that $\Omega_{\rm DM} h^2$ is almost insensitive to $g_X$ as $M_-$ increase along the red line segment,
 as opposed to the naive expectation which dictates the increase of $g_X$ to compensate for the increase of $M_-$. 
This is due to enhancement $M_-^4/m_{Z'}^4$ of the
$N_- N_- \to Z' Z'$ cross section which comes from the longitudinal component of $Z'$~\cite{Baek:2015fea}.
Since $C_9^{\mu,{\rm NP}}$ mildly depends on $M_-$ and does not depend on $g_X$, wide region can explain the $b \to s \mu\mu$ anomaly
at 1$\sigma$ level, accommodating the correct DM relic abundance at the same time.
As in the case of Scenario I) the direct detection occurs via effective operators of types in (\ref{eq:DD1}) and (\ref{eq:DD2}).
And the constraint from the direct detection experiments is not significant. 
Now let us discuss the fate of $Z'$ in this scenario. Since $Z'$ is not protected by the symmetry, it decays eventually
into the SM particles.
A main contribution comes from one-loop diagram where $N_\mp$ and $\wt{\ell}$ are running. A naive estimate for the partial decay width
\begin{align}
\Gamma(Z' \to \mu^- \mu^+) \sim m_{Z'} {g_X^2 \over 4 \pi} \left( {(\lambda_\ell^2)^2 \over 16 \pi^2} \right)^2
\sim 5.0 \times 10^{-3} \left(m_{Z'} \over 10~\textrm{GeV}\right)\left(g_X \over 0.5\right)^2
\left(\lambda_\ell^2 \over 5\right)^4 \, {\rm GeV},
\end{align}
gives the lifetime, $ \sim 10^{-22}$ sec, which is much shorter than the age of the universe.
If the annihilation cross section of $N_-N_- \to Z'Z'$ in the current universe is too large, the experiments measuring cosmic rays will
impose constraints on the parameter space.
To see this more clearly we obtained the expression of $\sigma v$ for $t$-channel $N_+$-exchanging  $N_-N_- \to Z'Z'$ process:
\begin{align}
\sigma v (N_-N_- \to Z'Z') = \frac{g_X^4(M_-^2-m_{Z'}^2)^{3/2}}{2 \pi M_-(M_-^2+M_+^2-m_{Z'}^2)^2} +O(v^2).
\end{align}
For $g_X=0.5$, $M_-=10$ GeV, $m_{Z'}=5$ GeV, and $M_+=1$ TeV, we get 
\begin{align}
\sigma v (N_-N_- \to Z'Z') = 3.0 \times 10^{-29} \, {\rm cm^3/s},
\end{align}
which is much smaller than the Fermi-LAT bound of about $ 3.0 \times 10^{-27} \, {\rm cm^3/s}$~\cite{Ackermann:2015zua}.
The annihilation cross section for $\sigma v(N_- N_- \to \mu^- \mu^+ (\nu_\mu \ol{\nu}_\mu))$ in the current universe is very small because it
is $p$-wave suppressed as can be seen from (\ref{eq:NN2ll}).

The entire region in both plots in figure~\ref{fig:Zp}
satisfies the constraint from the $B_s-\ol{B}_s$ mixing (\ref{eq:BsBs}).

\begin{figure}[ht]
\begin{center}
\includegraphics[width=.45\textwidth]{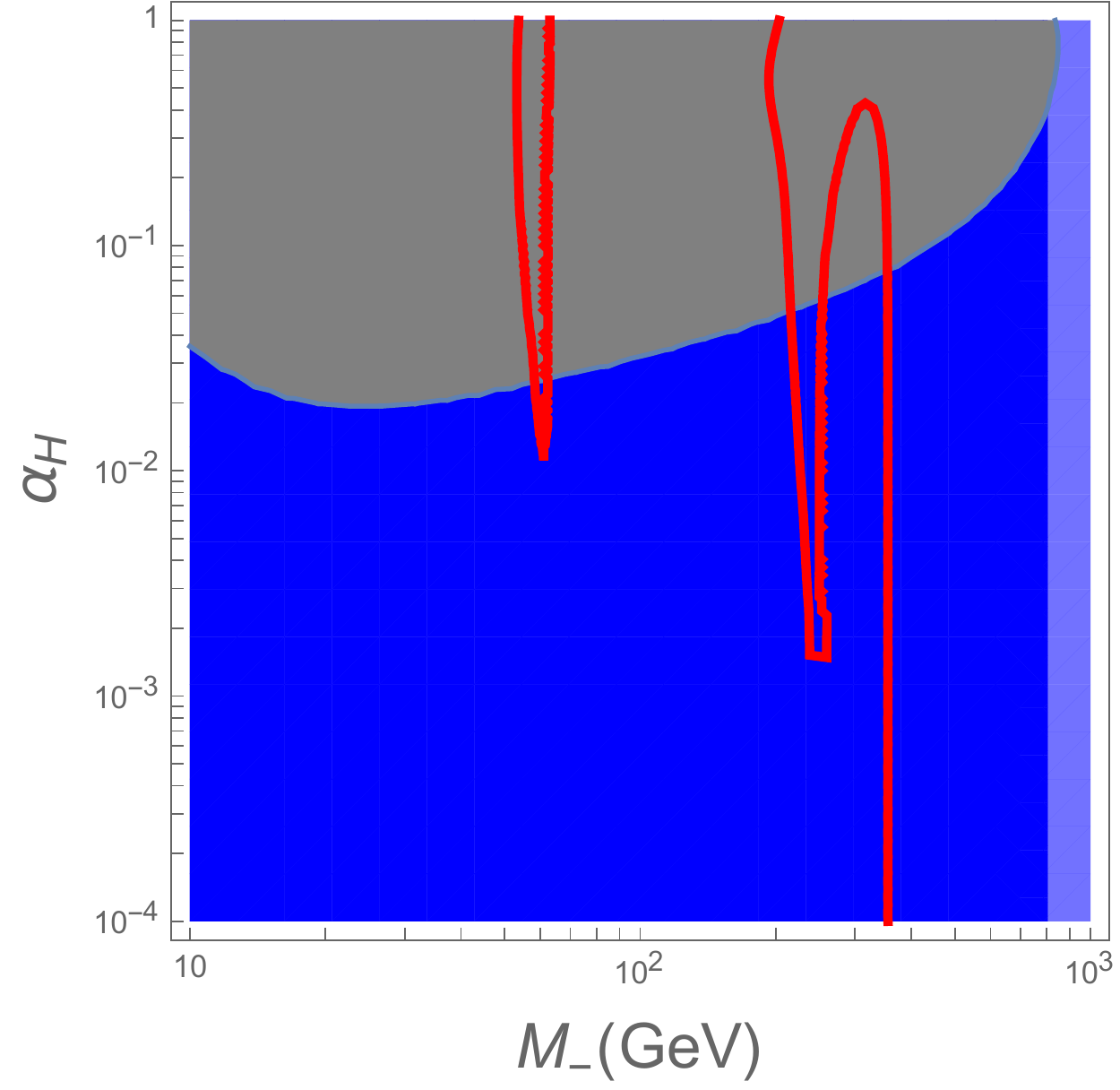}
\includegraphics[width=.45\textwidth]{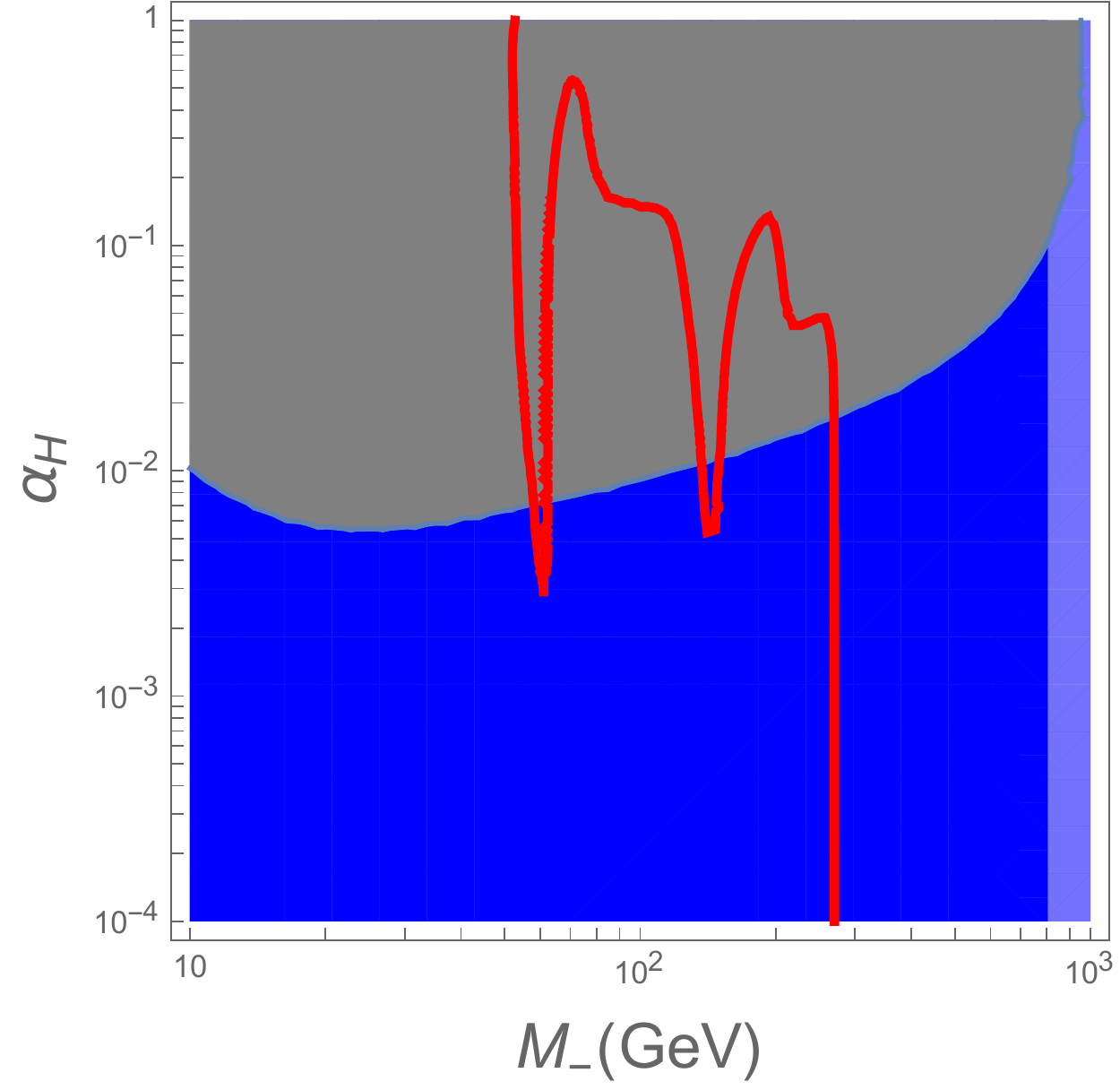}
\end{center}
\caption{DM relic density and $C_9^{\mu,{\rm NP}}$ in  $(M_-,\alpha_H)$-plane in Scenario III).   
The red line is a contour line for the constant dark matter relic density $\Omega_{\rm DM} h^2 = 0.12$.
The dark (medium) blue region satisfies $1 \sigma (2\sigma)$  allowed region for $C_9^{\mu,{\rm NP}}$ 
to explain $b \to s \mu\mu$ anomaly in (\ref{eq:C9_range}). The gray region is excluded by
a recent DM direct detection experiment XENON1T~\cite{Aprile:2018dbl}.
For the left (right) plot, we take $m_{H_2}=500 (300)$ GeV and $g_X = 0.5 (2)$.
The other parameters we take can be found in the text.}
\label{fig:HP}
\end{figure}

Now let us consider Scenario III) whose typical Feynman diagram for DM annihilation is figure~\ref{fig:DM_ann} III). 
A necessary condition that Scenario III) is dominant is to have sizable $\alpha_H$. 
We also need sizable $f$, which is not free parameter in our set up but is given by $f=\sqrt{2} g_X |Q| \Delta M/m_{Z'}$.
To enhance $f$ we need to enhance $g_X/m_{Z'}$, and as a consequence dark gauge interaction (Scenario II)) may interfere with this scenario.
Since we also need to explain the $b \to s\mu\mu$ anomaly, we need to allow  large $\lambda_\ell^2$ given that $\lambda_q^2 \lambda_q^{3*}$
is strongly constrained by $B_s-\ol{B}_s$ mixing. This also allows an enhancement of $N_- N_- \to \ell \ol{\ell}$
process through $t$-channel $\wt{\ell}$-exchange, which may result in the domination of Yukawa interaction $\lambda_\ell^2$ leading
to Scenario I) above. The above analysis tells us that {\it pure} Higgs portal scenario may not be possible in our model in case we want to
explain the $b \to s \mu\mu$ anomaly and dark matter at the same time.
The Higgs mixing angle $\alpha_H$ is also strongly constrained by the DM direct search experiments such as XENON1T, PANDA, {\it etc.}.
This can be seen in figure~\ref{fig:HP}. It shows DM relic density and $C_9^{\mu,{\rm NP}}$ in  $(M_-,\alpha_H)$-plane in Scenario III).   
The red line is a contour line for the constant dark matter relic density $\Omega_{\rm DM} h^2 = 0.12$.
The dark (medium) blue region satisfies $1 \sigma (2\sigma)$  allowed region for $C_9^{\mu,{\rm NP}}$ 
to explain $b \to s \mu\mu$ anomaly in (\ref{eq:C9_range}). The gray region is excluded by
a recent DM direct detection experiment XENON1T~\cite{Aprile:2018dbl}.
For the left (right) plot, we take $m_{H_2}=500 (300)$ GeV and $g_X = 0.5 (2)$.
For this figure we fixed as
\begin{align}
\lambda_{q}^2=-0.3, \quad  \lambda_{q}^3=0.5, \quad  \lambda_{\ell}^2=5, \quad m_{Z'} =M_+=1000 \, {\rm GeV}, 
\quad m_{\wt{q}}=m_{\wt{\ell}}=2000 \, {\rm GeV},
\end{align}
and (\ref{eq:scan_fixed}).
In the left (right) plot, for $M_- \lesssim 350 (250)$ GeV, the Higgs portal interaction can achieve the relic density but only in the
narrow resonance region near the SM Higgs $M_- \approx m_{H_1}/2 \approx 62.5$ GeV and the dark Higgs 
$M_- \approx m_{H_2}/2 \approx 250 (150)$ GeV.
When $M_- \gtrsim 350 (250)$ GeV, the $t$-channel $N_-$-exchanging process $N_- N_- \to H_2 H_2$ which is independent of 
$\alpha_H$  becomes important. This explains the abrupt drop of red line near the threshold.
As in the case of Scenario II), the $b \to s \mu\mu$ anomaly can be easily explained once the correct DM relic density is obtained.

\section{Conclusions}
\label{sec:concl}
We proposed a new physics model which can explain the recent $b \to s \mu\mu$ anomaly and a cold dark matter at the same time.
The model has a local dark $U(1)_X$ symmetry which is broken spontaneously into a discrete $Z_2$ symmetry by a dark Higgs scalar
$S$. This local discrete symmetry guarantees the stability of the dark matter. The dark matter candidates $N$ and the new 
$SU(2)_L$-doublet scalars $\wt{q}$ and $\wt{\ell}$ which have the same quantum numbers with the left-handed $SU(2)_L$-doublet quarks and
leptons contribute to $b \to s \mu\mu$ process via box diagrams.

We considered possible constraints on the model, which include $B_s-\ol{B}_s$ mixing, $B \to K^{(*)} \nu\ol{\nu}$ decay, 
inclusive $B$-decay $b \to s\gamma$, $B_s\to \mu^+ \mu^-$, anomalous magnetic moment of muon $a_\mu$, loop-induced effective $Z\mu^+\mu^-$ vertex
as well as new particle masses from the LHC.
We also checked whether the correct dark matter relic abundance can be achieved with the constraint from the dark matter direct detection
experiments.
We found that the constraint from $B_s-\ol{B}_s$ mixing is the strongest in the flavor sector.
The $b \to s\mu\mu$ anomaly can be explained by assuming a relatively large $\lambda_{\ell}^2 \approx 2$ for
TeV new particles while satisfying the $B_s -\ol{B}_s$ mixing constraint with $\lambda_{q}^2 \lambda_{q}^3 \approx -0.15$.

Given the large Yukawa coupling $\lambda_{\ell}^2 \approx 2$, the $t$-channel $N_- N_- \to \ell \ol{\ell}$ plays an important role in
achieving  the current relic abundance of the universe, showing a strong interplay between apparently unrelated flavor and dark
matter physics. When $Z'$ is lighter than the dark matter and the dark gauge coupling $g_X$ is sizable, $N_- N_- \to Z' Z'$ can also become
dominant. These two dark matter annihilation processes are not strongly constrained by the dark matter direct detection experiments because
the dark matter scattering with the nucleon processes occur first at one-loop level and are suppressed by the dark matter Yukawa coupling
with the first generation quarks. On the other hand the Higgs portal interaction can play important role to generate the current dark matter
relic only near the resonance region and is strongly constrained by the direct detection experiments, restricting the mixing angle of the SM
Higgs and dark Higgs $\alpha_H \lesssim 0.01$. 

\appendix
\section{Loop functions}
\label{app:loop}
The loop functions with multiple arguments for box diagrams of $b \to s \mu\mu$ and $B_s-\ol{B}_s$ mixing are defined recursively as
\begin{align}
f(x_1,x_2, x_3, \cdots) \equiv \frac{f(x_1,x_3, \cdots) - f(x_2, x_3, \cdots)}{x_1 - x_2}, 
\end{align}
where $f=j,k$ given by
\begin{align}
j(x) &= \frac{ x \log x}{x-1}, \nl
k(x) &= \frac{ x^2 \log x}{x-1}. 
\end{align}
For example, 
\begin{align}
j(x,y) &= \frac{ j(x)-j(y) }{x-y} = \frac{(y-1) x\log x-(x-1) y \log y }{(x-y)(x-1)(y-1)}.
\end{align}
We get $k(1,1,1)=1/3$ and $j(1,1,1)=-1/6$.
The loop function for $\gamma$-penguin is
\begin{align}
P_\gamma(x)=\frac{2-9x+18x^2-11x^3+6x^3 \log x}{36(1-x)^4}.
\end{align}
The loop function $J_1(y)$ for $b \to s \gamma$ is obtained to be
\begin{align}
J_1(y) &=\frac{1-6y+3y^2+2y^3-6y^2 \log y}{12(1-y)^4}.
\end{align}
We have $J_1(1)=1/24$. The loop function for the effective $Z\mu\mu$ vertex $\wt{F}_9(y)$ is obtained as an approximate analytic form of more general
Passarino-Veltman one-loop integrals:
\begin{align}
2 C_{00}(0,q^2,0,m^2_\psi,m^2_\phi,m^2_\phi)-B_0(0,m^2_\phi,m^2_\psi)-B_1(0,m^2_\phi,m^2_\psi) \approx -\frac{q^2}{m^2_\phi}\wt{F}_9(y),
\end{align}
where $y=m^2_\psi/m^2_\phi$ and
\begin{align}
\wt{F}_9(y) =\frac{-2+9y-18y^2+11y^3-6y^3 \log y}{36(1-y)^4}.
\end{align}
As a special case we get $\wt{F}_9(1)=-1/24$.

\acknowledgments
This work was supported in part by the National Research Foundation of Korea(NRF) grant funded by
     the Korea government(MSIT), Grant No. NRF-2018R1A2A3075605 (S.B.), 
     NRF-2017R1A2B4011946 (C.Y.), and NRF-2017R1E1A1A01074699 (C.Y.) and in part by a Korea University Grant (C.Y.)

\bibliographystyle{JHEP}
\bibliography{RK_box}

\end{document}